\documentclass[11pt]{article}

\usepackage{research5}
\usepackage{epsfig}
\usepackage{psfrag}
\usepackage{multicol}
\usepackage{amsmath}
\usepackage{amsthm}
\usepackage{IEEEtrantools}
\usepackage{verbatim}

\theoremstyle{plain}
\newtheorem{thm}{Theorem}[section]

\newtheorem{cor}{Corollary}[section]
\newtheorem{lm}{Lemma}[section]
\theoremstyle{definition}
\newtheorem{dfn}{Definition}[section]
\newtheorem{assumption}{Assumption}[section]
\newtheorem{rdc}{Reduction}[section]
\theoremstyle{remark}
\newtheorem{rmk}{Remark}[section]

\title{Sending a Bivariate Gaussian Source over a\\ Gaussian MAC with
  Feedback}

\author{Amos Lapidoth \and Stephan Tinguely}

\date{}

\begin{document}

\maketitle

\begin{abstract}
\renewcommand{\thefootnote}{}

We study the power-versus-distortion trade-off for the transmission of
a memoryless bivariate Gaussian source over a two-to-one Gaussian
multiple-access channel with perfect causal feedback. In this problem,
each of two separate transmitters observes a different component of a
memoryless bivariate Gaussian source as well as the feedback from the
channel output of the previous time-instants. Based on the observed
source sequence and the feedback, each transmitter then describes its
source component to the common receiver via an average-power
constrained Gaussian multiple-access channel. From the resulting
channel output, the receiver wishes to reconstruct both source
components with the least possible expected squared-error
distortion. We study the set of distortion pairs that can be achieved
by the receiver on the two source components.

We present sufficient conditions and necessary conditions for the
achievability of a distortion pair. These conditions are expressed in
terms of the source correlation and of the signal-to-noise ratio (SNR)
of the channel. In several cases the necessary conditions and
sufficient conditions coincide. This allows us to show that if the
channel SNR is below a certain threshold, then an uncoded transmission
scheme that ignores the feedback is optimal. Thus, below this
SNR-threshold feedback is useless. We also derive the precise high-SNR
asymptotics of optimal schemes.



\footnote{The work of Stephan Tinguely was partially supported by the
  Swiss National Science Foundation under Grant 200021-111863/1. The
  results in this paper were presented in part at the 2007 IEEE
  International Symposium on Information Theory, Nice, France.

  A.~Lapidoth and S.~Tinguely are with the Signal and Information
  Processing Laboratory (ISI), ETH Zurich, Switzerland (e-mail:
  lapidoth@isi.ee.ethz.ch; tinguely@isi.ee.ethz.ch).}
\setcounter{footnote}{0}
\end{abstract}

\section{Introduction}

This is a sequel to the work in \cite{lapidoth-tinguely08-mac-it}
where a bivariate Gaussian source is to be transmitted over a
Gaussian multiple-access channel. The new element here is the presence
of perfect causal feedback from the channel output to each of the
transmitters. As in \cite{lapidoth-tinguely08-mac-it}, our interest is
in the power-versus-distortion trade-off.

Our setup consists of a memoryless bivariate Gaussian source and a
two-to-one Gaussian multiple-access channel with perfect causal
feedback. Each of the two transmitters in the multiple-access channel
observes a different component of the source as well as feedback from
the previous channel outputs. Based on the feedback and the observed
source sequence, each transmitter then describes its source component
to the common receiver via an average-power constrained Gaussian
multiple-access channel. From the output of the channel, the receiver
wishes to reconstruct both source components with the least possible
expected squared-error distortion. Our interest is in characterizing
the pairs of squared-error distortions that can be achieved
simultaneously on the two source components.

We present sufficient conditions and necessary conditions for the
achievability of a distortion pair. These conditions are expressed in
terms of the source correlation and the signal-to-noise ratio (SNR) of
the channel. In several cases the necessary conditions and sufficient
conditions are shown to agree. In particular, we show that if the
channel SNR is below a certain threshold, then an uncoded transmission
scheme is optimal, and feedback is useless. We also show that, in
general the source-channel separation approach is suboptimal, but that
it is asymptotically optimal as the transmit power tends to infinity.









\section{Problem Statement}\label{sec:problem-statement}

\subsection{Setup}\label{subsec:setup}

\begin{figure}[h]
  \centering
  \psfrag{s1}[cc][cc]{$S_{1,k}$}
  \psfrag{s2}[cc][cc]{$S_{2,k}$}
  \psfrag{x1}[cc][cc]{$X_{1,k}$}
  \psfrag{x2}[cc][cc]{$X_{2,k}$}
  \psfrag{src}[cc][cc]{Source}
  \psfrag{f1}[cc][cc]{$f_{1,k}^{(n)}(\cdot)$}
  \psfrag{f2}[cc][cc]{$f_{2,k}^{(n)}(\cdot)$}
  \psfrag{z}[cc][cc]{$Z_{k}$}
  \psfrag{y}[cc][cc]{$Y_{k}$}
  \psfrag{p1}[cc][cc]{$\phi_1^{(n)}(\cdot)$}
  \psfrag{p2}[cc][cc]{$\phi_2^{(n)}(\cdot)$}
  \psfrag{s1h}[cc][cc]{$\hat{S}_{1,k}$}
  \psfrag{s2h}[cc][cc]{$\hat{S}_{2,k}$}
  \epsfig{file=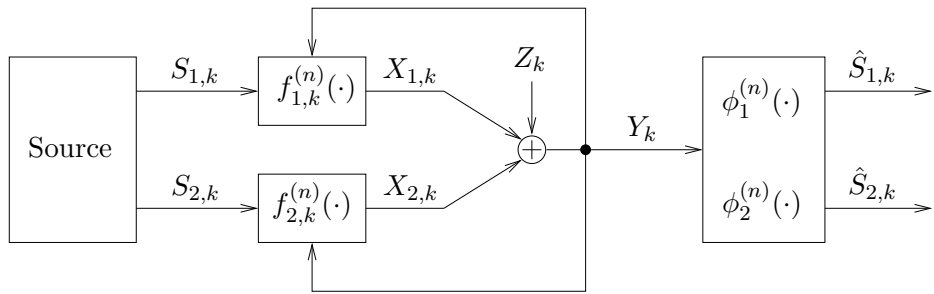, width=0.85\textwidth}
  \caption{Bivariate Gaussian source with one-to-two Gaussian
    multiple-access channel with feedback.}
  \label{fig:setup-macfb}
\end{figure}

Our setup is illustrated in Figure \ref{fig:setup-macfb}.
A memoryless bivariate Gaussian source is connected to a two-to-one
Gaussian multiple-access channel with perfect causal feedback. Each
transmitter of the multiple-access channel observes one of the source
components and wishes to describe it to the common receiver. The
source symbols produced at time $k \in \Integers$ are denoted by
$(S_{1,k}, S_{2,k})$. The source output pairs $\{ (S_{1,k}, S_{2,k})
\}$ are independent identically distributed (IID) zero-mean Gaussians
of covariance matrix
\begin{equation}\label{eq:source-law}
\cov{S} = \left( \hspace{-0.8mm} \begin{array}{c c}
\sigma_1^2 & \hspace{-1mm} \rho \sigma_1 \sigma_2\\[4mm]
\rho \sigma_1 \sigma_2 & \hspace{-1mm} \sigma_2^2
\end{array} \hspace{-1.5mm} \right),
\end{equation}
where $\rho \in [-1,1]$ and $0 < \sigma_i^2 < \infty$, $i \in \{ 1,2 \}$. The
sequence $\{ S_{1,k} \}$ of the first source component is observed by
Transmitter~1 and the sequence $\{ S_{2,k} \}$ of the second source
component is observed by Transmitter~2. The two source components are
to be described over the multiple-access channel to the common
receiver by means of the channel input sequences $\{ X_{1,k} \}$ and
$\{ X_{2,k} \}$, where $x_{1,k} \in \Reals$ and $x_{2,k} \in
\Reals$. The corresponding time-$k$ channel output is given by
\begin{equation}
Y_k = X_{1,k} + X_{2,k} + Z_k,
\end{equation}
where $Z_k$ is the time-$k$ additive noise term, and where $\{ Z_{k}\}$
are IID zero-mean variance-$N$ Gaussian random variables that are
independent of the source sequence.

We consider block encoding schemes and denote the block-length by $n$
and the associated $n$-sequences in boldface, e.g.~${\bf S}_1 =
(S_{1,1}, S_{1,2}, \ldots , S_{1,n})$. Transmitter~$i~\in~\{ 1,2 \}$
is described by a sequence of functions $f_{i,k}^{(n)} \colon
\Reals^{n} \times \Reals^{k-1} \rightarrow \Reals$, $k = 1,\ldots ,n$,
which, for every time instant $k \in \Reals$ produce the channel input
$X_{i,k}$ from the source sequence ${\bf S}_i$ and the so-far-observed
feedback sequence $Y^{k-1} = (Y_1, \ldots ,Y_{k-1})$, i.e.
\begin{IEEEeqnarray}{rCl}\label{eq:encoders-fb}
X_{i,k} & = & f_{i,k}^{(n)} \left( {\bf S}_i, Y^{k-1} \right) \qquad i
\in \{ 1,2 \}.
\end{IEEEeqnarray}
The channel input sequences are subjected to expected average power
constraints
\begin{IEEEeqnarray}{rCl}\label{eq:power-constraint}
  \frac{1}{n} \sum_{k=1}^n \E{X_{i,k}^2} & \leq & P_i \qquad i \in \{
  1,2 \},
\end{IEEEeqnarray}
for some given $P_i > 0$.

The receiver is described by two functions $\phi_{i}^{(n)} \colon
\Reals^{n} \rightarrow \Reals^{n}$, $i \in \{ 1,2 \}$, each of which
forms an estimate $\hat{\bf S}_i$ of the respective source sequence
${\bf S}_i$ based on the observed channel output sequence ${\bf
  Y}$. Thus,
\begin{IEEEeqnarray}{rCl}\label{eq:reconstructors}
  \hat{\bf S}_i & = & \phi_{i}^{(n)} \left( {\bf Y} \right) \qquad i
  \in \{ 1,2 \}.
\end{IEEEeqnarray}

We are interested in the pairs of expected squared-error distortions
that can be achieved simultaneously on the source-pair as the
blocklength $n$ tends to infinity. In view of this, we next define the
notion of achievability.

\subsection{Achievability of Distortion Pairs}\label{subsec:achievability}

\begin{dfn}\label{def:achv-dist}
  Given $\sigma_{1}, \sigma_{2} > 0$, $\rho \in [-1,1]$, $P_{1}, P_{2}
  > 0$, and $N > 0$ we say that the tuple $(D_{1}, D_{2},
  \sigma^{2}_{1}, \sigma_{2}^{2},$ $\rho, P_{1}, P_{2}, N )$ is
  \emph{achievable} if there exists a sequence of encoding functions
  $\big( \{ f_{1,k}^{(n)} \}_{k=1}^n, \{ f_{2,k}^{(n)} \}_{k=1}^n
  \big)$ as in \eqref{eq:encoders-fb}, satisfying the average power
  constraints (\ref{eq:power-constraint}), and a sequence of
  reconstruction pairs $\big( \phi_{1}^{(n)}, \phi_{2}^{(n)} \big)$ as in
  (\ref{eq:reconstructors}), such that the average distortions
  resulting from these encoding and reconstruction functions fulfill
  \begin{displaymath}
    \varlimsup_{n \rightarrow \infty} \frac{1}{n} \sum_{k=1}^n \E{
      \left( S_{i,k} - \hat{S}_{i,k} \right)^2} \leq D_{i}, \quad i
    \in \{ 1,2 \},
  \end{displaymath}
  whenever
  \begin{displaymath}
    Y_k = f_{1,k}^{(n)}({\bf S}_1,Y^{k-1}) + f_{2,k}^{(n)}({\bf S}_2,Y^{k-1}) +
    Z_k, \qquad \text{for } k \in \{ 1,2, \ldots ,n \},
  \end{displaymath}
  and where $\{(S_{1,k},S_{2,k})\}$ are IID zero-mean bivariate
  Gaussian vectors of covariance matrix $\cov{S}$ as in
  \eqref{eq:source-law} and $\{Z_{k}\}$ are IID zero-mean variance-$N$
  Gaussians that are independent of $\{(S_{1,k},S_{2,k})\}$.
\end{dfn}

For given $\sigma_1^{2}$, $\sigma_2^{2}$, $\rho$, $P_{1}$, $P_{2}$,
and $N$, we wish to find the set of pairs $(D_{1}, D_{2})$ such that
$(D_{1}, D_{2}, \sigma_1^{2}, \sigma_2^{2}, \rho, P_{1}, P_{2}, N)$ is
achievable. Sometimes, we will refer to the set of all $(D_1,D_2)$
such that $(D_{1}, D_{2}, \sigma_1^{2}, \sigma_2^{2}, \rho, P_{1},
P_{2}, N)$ is achievable as the distortion region associated with $(
\sigma_1^{2}, \sigma_2^{2}, \rho, P_{1}, P_{2}, N )$. In that sense,
we will often say, with respect to some $( \sigma_{1}$, $\sigma_{2}$,
$\rho$, $P_{1}$, $P_{2}, N )$, that the pair $(D_1,D_2)$ is
achievable, instead of saying that the tuple $(D_{1}, D_{2},
\sigma^{2}_{1}, \sigma_{2}^{2},$ $\rho, P_{1}, P_{2}, N )$ is
achievable.

\subsection{Normalization}

For the described problem we now show that, without loss in
generality, the source law given in \eqref{eq:source-law} can be
restricted to a simpler form. This restriction will ease the statement
of our results as well as their derivations.

\begin{rdc}\label{rdc:source-normalization}
  For the problem stated in Sections \ref{subsec:setup} and
  \ref{subsec:achievability}, there is no loss in generality in
  restricting the source law to satisfy
  \begin{IEEEeqnarray}{rCl}\label{eq:source-normalization}
    \sigma_1^2 = \sigma_2^2 = \sigma^2 & \hspace{12mm} \text{and}
    \hspace{12mm} & \rho \in [0,1].
  \end{IEEEeqnarray}
\end{rdc}

\begin{proof}
  The proof follows by noting that the described problem has certain
  symmetry properties with respect to the source law. We prove the
  reductions on the source variance and on the correlation coefficient
  separately.
  \renewcommand{\labelenumi}{\roman{enumi})}
  \begin{enumerate}
  \item The reduction to correlation coefficients $\rho \in [0,1]$
    holds because the optimal distortion region depends on the
    correlation coefficient only via its absolute value $|\rho
    |$. That is, the tuple $(D_{1}, D_{2}, \sigma_{1}^{2},
    \sigma_{2}^{2}, \rho,$ $P_{1}, P_{2},N)$ is achievable if, and
    only if, the tuple $(D_{1}, D_{2}, \sigma_{1}^{2}, \sigma_{2}^{2},
    -\rho, P_{1}, P_{2},N)$ is achievable.  To see this, note that if
    $\big( \{ f_{1,k}^{(n)} \}_{k=1}^n, \{ f_{2,k}^{(n)} \}_{k=1}^n,
    \phi_{1}^{(n)}, \phi_{2}^{(n)} \big)$ achieves the distortion
    $(D_{1}, D_{2})$ for the source of correlation coefficient $\rho$,
    then $\big( \{ \tilde{f}_{1,k}^{(n)} \}_{k=1}^n, \{ f_{2,k}^{(n)}
    \}_{k=1}^n, \tilde{\phi}_{1}^{(n)}, \phi_{2}^{(n)} \big)$, where
    \begin{equation*}
      \tilde{f}_{1,k}^{(n)}( {\bf S}_{1}, Y^{k-1} ) = f_{1,k}^{(n)}( -
      {\bf S}_{1}, Y^{k-1} ) \qquad \text{and} \qquad 
      \tilde{\phi}_{1}^{(n)}( {\bf Y} ) = - \phi_{1}^{(n)}( {\bf Y} ) 
    \end{equation*}
    achieves  $(D_{1}, D_{2})$ on the source with correlation
    coefficient $-\rho$.
    
  \item The restriction to source variances satisfying $\sigma_1^2 =
    \sigma_2^2 = \sigma^2$ incurs no loss of generality because the
    distortion region scales linearly with the source variances. That
    is, the tuple $(D_{1}, D_{2}, \sigma_{1}^{2}, \sigma_{2}^{2},
    \rho, P_{1}, P_{2},N)$ is achievable if, and only if, for every
    $\alpha_{1}, \alpha_{2} \in \Reals^+$, the tuple
    $(\alpha_{1}D_{1}, \alpha_{2}D_{2}, \alpha_{1} \sigma_{1}^{2},
    \alpha_{2} \sigma_{2}^{2}, \rho, P_{1}, P_{2},N)$ is achievable.

    This can be seen as follows. If $\big( \{f_{1,k}^{(n)}\}_{k=1}^n,
    \{f_{2,k}^{(n)}\}_{k=1}^n, \phi_{1}^{(n)}, \phi_{2}^{(n)} \big)$
    achieves the tuple $(D_{1}, D_{2}, \sigma_{1}^{2}, \sigma_{2}^{2}, \rho,
    P_{1}, P_{2},N)$, then the combination of the encoders 
    \begin{equation*}
      \tilde{f}_{i,k}^{(n)}( {\bf S}_{i}, Y^{k-1}) = f_{i,k}^{(n)}( {\bf
        S}_{i}/\sqrt{\alpha_{i}}, Y^{k-1}), \hspace{10mm} i \in \{ 1,2 \},
    \end{equation*}
    with the reconstructors
    \begin{equation*}
      \tilde{\phi}_{i}^{(n)}( {\bf Y} ) = \sqrt{\alpha_{i}} \cdot
      \phi_{i}^{(n)}( {\bf Y} ), \hspace{10mm} i \in \{ 1,2 \},
    \end{equation*}
    achieves the tuple $(\alpha_1 D_{1}, \alpha_2 D_{2}, \alpha_{1}
    \sigma_{1}^{2}, \alpha_{2} \sigma_{2}^{2}, \rho, P_{1},
    P_{2},N)$. And by an analogous argument it follows that if
    $(\alpha_1 D_{1}, \alpha_2 D_{2}, \alpha_{1} \sigma_{1}^{2},
    \alpha_{2} \sigma_{2}^{2}, \rho, P_{1}, P_{2},N)$ is achievable,
    then also $(D_{1}, D_{2}, \sigma_{1}^{2}, \sigma_{2}^{2}, \rho,
    P_{1}, P_{2},N)$ is achievable. \hfill \qedhere
  \end{enumerate}
\end{proof}

In view of Reduction \ref{rdc:source-normalization} we assume for the
remainder that the source law additionally satisfies
\eqref{eq:source-normalization}.

\subsection{``Symmetric Version'' and a Convexity Property}\label{subsec:symmetry}

The ``symmetric version'' of our problem corresponds to the case where
the transmitters are subjected to the same power constraint, and where
we seek to achieve the same distortion on each source component. That
is, $P_{1} = P_{2} = P$, and we are interested in the minimal
distortion
\begin{IEEEeqnarray*}{rCl}
  D^{*}(\sigma^{2}, \rho, P, N) & \triangleq & \inf \{ D \colon (D, D,
  \sigma^{2}, \sigma^{2}, \rho, P, P, N) \text{ is achievable}\},
\end{IEEEeqnarray*}
that is simultaneously achievable on $\{ S_{1,k} \}$ and on $\{
S_{2,k} \}$. In this case, we will often express the distortion
$D^{*}(\sigma^{2}, \rho, P, N)$, for some fixed $\sigma^2$ and $\rho$,
and as a function of the SNR $P/N$.

We conclude this section with a convexity property of the achievable
distortions.

\begin{rmk}\label{rmk:convexity}
  If $(D_{1}, D_{2}, \sigma_{1}^{2}$, $\sigma_{2}^{2}, \rho, P_{1},
  P_{2}, N)$ and $(\tilde{D}_{1}, \tilde{D}_{2}, \sigma_{1}^{2},
  \sigma_{2}^{2}, \rho, \tilde{P}_{1}, \tilde{P}_{2}, N)$ are
  achievable, then
  \begin{equation*}
    \left( \lambda D_{1} + \bar{\lambda} \tilde{D}_{1}, \lambda D_{2} + 
      \bar{\lambda}\tilde{D}_{2},  \sigma_{1}^{2}, \sigma_{2}^{2}, \rho,
      \lambda P_{1} + \bar{\lambda} \tilde{P}_{1}, \lambda P_{2} + 
      \bar{\lambda}\tilde{P}_{2}, N \right),
  \end{equation*}
  is also achievable for every $\lambda \in [0,1]$, where
  $\bar{\lambda} = (1-\lambda)$.
\end{rmk}

\begin{proof}
  Follows by a time-sharing argument.
\end{proof}



\section{Main Results}

\subsection{Necessary Condition for Achievability of $(D_1,D_2)$}

To state our necessary condition we first introduce three
rate-distortion functions. They are: the rate-distortion function
$R_{S_1,S_2}(D_1,D_2)$ on $\{ (S_{1,k}, S_{2,k}) \}$; the
rate-distortion function $R_{S_1|S_2}(D_1)$ on $\{ S_{1,k} \}$, when
the component $\{ S_{2,k} \}$ is observed as side-information at both,
encoder and decoder; and the rate-distortion function
$R_{S_2|S_1}(D_2)$ on $\{ S_{2,k} \}$ when the component $\{ S_{1,k}
\}$ is observed as side-information at both, encoder and decoder. For
$\{ (S_{1,k},S_{2,k}) \}$ jointly Gaussian as in \eqref{eq:source-law}
with $\sigma_1^2 = \sigma_2^2 = \sigma^2$, the two latter functions are
given by
\begin{IEEEeqnarray}{rCl}
  R_{S_1|S_2}(D_1) & = & \frac{1}{2} \log_2^+ \left(
    \frac{\sigma^2(1-\rho^2)}{D_1}\right), \label{eq:Rd-S1|S2} \\[3mm]
  R_{S_2|S_1}(D_2) & = & \frac{1}{2} \log_2^+ \left(
    \frac{\sigma^2(1-\rho^2)}{D_2}\right). \label{eq:Rd-S2|S1}
\end{IEEEeqnarray}
The function $R_{S_1,S_2}(D_1,D_2)$ is given in the following theorem.

\begin{thm}[Xiao, Luo \cite{xiao-luo05}; Lapidoth, Tinguely
  \cite{lapidoth-tinguely08-mac-it, lapidoth-tinguely06}] \label{thm:rd1d2-main}
  The rate-distortion function $R_{S_1,S_2}(D_1,D_2)$ is given by
  \begin{IEEEeqnarray}{rCl}\label{eq:RD1D2-solution}
    R_{S_1,S_2}(D_1,D_2) & = & \left\{ \begin{array}{l c}
        \frac{1}{2} \log_2^+ \left( \frac{\sigma^2}{D_\textnormal{min}}
        \right) & \text{if } (D_1,D_2) \in \mathscr{D}_1\\[5mm]
        \frac{1}{2} \log_2^+ \left( \frac{\sigma^4 (1-\rho^2)}{D_1 D_2} \right)
        & \text{if } (D_1,D_2) \in \mathscr{D}_2\\[5mm]
        \frac{1}{2} \log_2^+ \left( \frac{\sigma^4 (1-\rho^2)}{D_1D_2 - \left(
              \rho \sigma^2 - \sqrt{(\sigma^2-D_1)(\sigma^2-D_2)} \right)^2}
        \right) & \text{if } (D_1,D_2) \in \mathscr{D}_3.
      \end{array} \right. \hspace{4mm}
  \end{IEEEeqnarray}
  where $\log_2^+(x) = \max \{ 0,\log_2(x) \}$, $D_{\textnormal{min}}
  = \min \left\{ D_1,D_2 \right\}$ and where the regions
  $\mathscr{D}_1$, $\mathscr{D}_2$ and $\mathscr{D}_3$ are given by
  \begin{IEEEeqnarray*}{rCl}
    \mathscr{D}_1 & = \Bigg\{ (D_1,D_2): \: &0 \leq D_1 \leq \sigma^2
    (1-\rho^2), \: \, D_2 \geq \sigma^2 (1-\rho^2) + \rho^2 D_1;\\
    & &\sigma^2(1-\rho^2) < D_1 \leq \sigma^2, \: \, D_2 \geq \sigma^2
    (1-\rho^2) + \rho^2 D_1,\\
    & & \hspace{73mm} D_2 \leq \frac{D_1 - \sigma^2(1-\rho^2)}{\rho^2}
    \Bigg\},\\[3mm]
    \mathscr{D}_2 & = \bigg\{ (D_1,D_2): \; &0 \leq D_1 \leq \sigma^2
    (1-\rho^2), 0 \leq D_2 < (\sigma^2(1-\rho^2) - D_1)
    \frac{\sigma^2}{\sigma^2-D_1} \bigg\},\\[3mm]
    \mathscr{D}_3 & = \Bigg\{ (D_1,D_2): \: &0 \leq D_1 \leq
    \sigma^2(1-\rho^2),\\
    & & \hspace{13mm} (\sigma^2(1-\rho^2) - D_1)
    \frac{\sigma^2}{\sigma^2-D_1} \leq D_2 < \sigma^2 (1-\rho^2) +
    \rho^2 D_1;\\[4mm]
    & & \hspace{-2mm} \sigma^2(1-\rho^2) < D_1 \leq \sigma^2, \: \, \frac{D_1 -
      \sigma^2(1-\rho^2)}{\rho^2} < D_2 < \sigma^2 (1-\rho^2) + \rho^2
    D_1 \Bigg\}.
  \end{IEEEeqnarray*}
\end{thm}

Our necessary condition is now as follows.

\begin{thm}\label{thm:general-bound}
  A necessary condition for the achievability of
  $(D_1,D_2,\sigma^2,\sigma^2,\rho,P_1,P_2, N)$ is the existance of
  some $\hat{\rho} \in [0,1]$ such that
  \begin{IEEEeqnarray}{rCl}
    R_{S_1,S_2}(D_1,D_2) & \leq & \frac{1}{2} \log_2 \left( 1 +
      \frac{P_1+P_2+2\hat{\rho} \sqrt{P_1P_2}}{N}\right) \label{eq:thm-cond1}\\[2mm]
    R_{S_1|S_2}(D_1) & \leq & \frac{1}{2} \log_2 \left( 1 +
      \frac{P_1(1-\hat{\rho}^2)}{N} \right) \label{eq:thm-cond2}\\[2mm]
    R_{S_2|S_1}(D_2) & \leq & \frac{1}{2} \log_2 \left( 1 +
      \frac{P_2(1-\hat{\rho}^2)}{N} \right). \label{eq:thm-cond3}
  \end{IEEEeqnarray}
\end{thm}

\begin{proof}
See Appendix \ref{appx:prf-thm-lbFB}.
\end{proof}

We now specialize Theorem \ref{thm:general-bound} to the symmetric
case. To this end, we first substitute the rate-distortion functions
$R_{S_1,S_2}(D_1,D_2)$, $R_{S_1 | S_2}(D_1)$, $R_{S_2 | S_1}(D_2)$ on
the LHS of \eqref{eq:thm-cond1} -- \eqref{eq:thm-cond3} by their
explicit forms given in \eqref{eq:RD1D2-solution},
\eqref{eq:Rd-S1|S2}, and \eqref{eq:Rd-S2|S1}
respectively. Substituting $(D,D)$ for $(D_1,D_2)$ in
\eqref{eq:thm-cond1} \& \eqref{eq:RD1D2-solution} yields that if
$(D,D)$ is achievable, then
\begin{IEEEeqnarray}{rCl}\label{eq:sym-cond1}
D & \geq & \left\{ \begin{array}{l l}
\frac{1}{2} \left( \frac{N \sigma^2 (1+\rho)}{N+2P(1+\hat{\rho})} +
  \sigma^2(1-\rho)\right) & \text{if }\frac{P}{N} \leq
\frac{\rho}{1-\rho^2}\\[3mm]
\sigma^2 \sqrt{ \frac{N (1-\rho^2)}{N+2P(1+\hat{\rho})}} & \text{if
}\frac{P}{N} > \frac{\rho}{1-\rho^2}.
\end{array} \right.
\end{IEEEeqnarray}
Similarly, from \eqref{eq:thm-cond2} \& \eqref{eq:Rd-S1|S2} [or
\eqref{eq:thm-cond3} \& \eqref{eq:Rd-S2|S1}] we obtain that if $(D,D)$
is achievable, then
\begin{IEEEeqnarray}{rCl}\label{eq:sym-cond2}
  D & \geq & \sigma^2 \frac{N (1-\rho^2)}{N+P(1-\hat{\rho}^2)}.
\end{IEEEeqnarray}
Denoting the RHS of \eqref{eq:sym-cond1} by $\xi (\sigma^2, \rho, P,
N, \hat{\rho})$ and the RHS of \eqref{eq:sym-cond2} by $\psi
(\sigma^2, \rho, P, N, \hat{\rho})$, gives the following lower bound on $D^{\ast}
(\sigma^2, \rho, P, N)$:

\begin{cor}\label{cor:symmetric-case}
  In the symmetric case
  \begin{IEEEeqnarray*}{rCl}
    D^{\ast}(\sigma^2,\rho,P,N) & \geq & \min_{0
      \leq \hat{\rho} \leq 1} \max \left\{ \xi (\sigma^2, \rho, P, N,
      \hat{\rho}), \psi (\sigma^2, \rho, P, N, \hat{\rho}) \right\}.
  \end{IEEEeqnarray*}
\end{cor}
The minimization over $\hat{\rho}$ is discussed in the following remark.

\begin{rmk}\label{rmk:lbFB-D*-opt}
  For $P/N \leq \rho^2/(2(1-\rho)(1+2\rho))$ the minimum in Corollary
  \ref{cor:symmetric-case} is achieved by $\hat{\rho}^{\ast} = 1$, and for
  all larger $P/N$ the minimum is achieved by the
  $\hat{\rho}^{\ast}$ satisfying
  \begin{IEEEeqnarray*}{rCl}
    \xi (\sigma^2, \rho, P, N, \hat{\rho}^{\ast}) & = & \psi (\sigma^2,
    \rho, P, N, \hat{\rho}^{\ast}).
  \end{IEEEeqnarray*}
  As $P/N \rightarrow \infty$ it can be shown that $\hat{\rho}^{\ast}$
  tends to one and hence Corollary \ref{cor:symmetric-case} yields
  \begin{IEEEeqnarray}{rCl}\label{eq:macfb-lb-hsnr}
    \varliminf_{P/N \rightarrow \infty} \sqrt{\frac{P}{N}}
    D^{\ast}(\sigma^2, \rho, P, N) & \geq & \sigma^2
    \sqrt{\frac{1-\rho^2}{4}}.
  \end{IEEEeqnarray}
\end{rmk}

In the next section we show that the $\liminf$ in
\eqref{eq:macfb-lb-hsnr} is a limit, and that it is achieved by
source-channel separation.

\subsection{Source-Channel Separation}

\begin{figure}[h]
  \centering
  \psfrag{s1}[cc][cc]{$S_{1,k}$}
  \psfrag{s2}[cc][cc]{$S_{2,k}$}
  \psfrag{r1}[cc][cc]{$R_1$}
  \psfrag{r2}[cc][cc]{$R_2$}
  \psfrag{src}[cc][cc]{Source}
  \psfrag{f1}[cc][cc]{$\bar{f}_1^{(n)}(\cdot)$}
  \psfrag{f2}[cc][cc]{$\bar{f}_2^{(n)}(\cdot)$}
  \psfrag{p1}[cc][cc]{$\bar{\phi}_1^{(n)}(\cdot)$}
  \psfrag{p2}[cc][cc]{$\bar{\phi}_2^{(n)}(\cdot)$}
  \psfrag{s1h}[cc][cc]{$\hat{S}_{1,k}$}
  \psfrag{s2h}[cc][cc]{$\hat{S}_{2,k}$}
  \epsfig{file=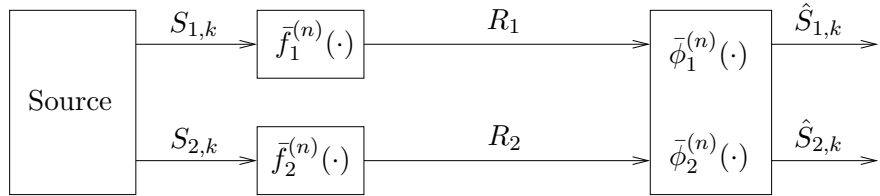, width=0.8\textwidth}
  \caption{Distributed source coding problem for a bivariate Gaussian
    source.}
  \label{fig:setup-oohama}
\end{figure}

We now consider the set of distortion pairs that are achieved by
combining the optimal scheme for the corresponding source-coding
problem with the optimal scheme for the corresponding channel-coding
problem. The source-coding problem is illustrated in Figure
\ref{fig:setup-oohama}.
The two source components are observed by two separate encoders. These
two encoders wish to describe their source sequence to the common
receiver by means of individual rate-limited and error-free bit
pipes. The receiver estimates each of the sequences subject to
expected squared-error distortion. A detailed description of this
problem can be found in \cite{oohama97,
  wagner-tavildar-vishwanath05}. The associated rate-distortion region
is given in the next theorem.
\begin{thm}[Oohama \cite{oohama97}; Wagner, Tavildar, and Viswanath
  \cite{wagner-tavildar-vishwanath05}]\label{thm:source-coding} 
  For the Gaussian two-terminal source coding problem (with source
  components of unit variances) a distortion-pair $(D_1,D_2)$ is
  achievable if, and only if,
  \begin{IEEEeqnarray*}{rCl}
    (R_1,R_2) \in \mathcal{R}_1(D_1) \cap \mathcal{R}_2(D_2) \cap
    \mathcal{R}_{\textnormal{sum}}(D_1,D_2),
  \end{IEEEeqnarray*} 
  where
  \begin{IEEEeqnarray*}{rCl}
    \mathcal{R}_1(D_1) & = & \left\{ (R_1,R_2): R_1\geq \frac{1}{2}
      \log_2^+ \left[ \frac{1}{D_1} (1-\rho^2(1-2^{-2R_2})) \right]
    \right\}\\[2mm]
    \mathcal{R}_2(D_2) & = & \left\{ (R_1,R_2): R_2\geq \frac{1}{2}
      \log_2^+ \left[ \frac{1}{D_2} (1-\rho^2(1-2^{-2R_1})) \right]
    \right\}\\[2mm]
    \mathcal{R}_{\textnormal{sum}}(D_1,D_2) & = & \left\{ (R_1,R_2):
      R_1+R_2\geq \frac{1}{2} \log_2^+ \left[ \frac{(1-\rho^2)
          \beta(D_1,D_2)}{2D_1D_2} \right] \right\}
  \end{IEEEeqnarray*}
  with
  \begin{IEEEeqnarray*}{rCl}
    \beta(D_1,D_2) & = & 1 + \sqrt{1 +
      \frac{4\rho^2D_1D_2}{(1-\rho^2)^2}}.
  \end{IEEEeqnarray*}
\end{thm}

The capacity region $\mathcal{C}_{FB}(P_1,P_2,N)$ of the Gaussian
multiple-access channel with feedback was derived in \cite{ozarow85}
and is restated in the following theorem.

\begin{thm}[Ozarow \cite{ozarow85}]\label{thm:capa-macfb}
  The capacity region $\mathcal{C}_{\text{FB}} (P_1,P_2,N)$ of the
  Gaussian multiple-access channel with perfect feedback is
  \begin{IEEEeqnarray*}{rCl}
    \mathcal{C}_{\text{FB}} (P_1,P_2,N) = \bigcup_{0 \leq \bar{\rho}
      \leq 1} \bigg\{ (R_1,R_2): 
    R_1 &\leq& \frac{1}{2} \log_2 \left( 1 +
      \frac{P_1(1-\bar{\rho}^2)}{N} \right) \\[2mm]
    R_2 &\leq& \frac{1}{2} \log_2 \left( 1 +
      \frac{P_2(1-\bar{\rho}^2)}{N} \right) \\[2mm]
    R_1+R_2 & \leq & \frac{1}{2} \log_2 \left( 1 +
      \frac{P_1+P_2+2\bar{\rho} \sqrt{P_1P_2}}{N}\right) \bigg\}. 
  \end{IEEEeqnarray*}
\end{thm}

The distortions achievable by source-channel separation are now given
in the following Corollary.

\begin{cor}\label{cor:FB-sep-based}
A distortion pair $(D_1,D_2)$ is achievable by source-channel
separation if, and only if,
\begin{IEEEeqnarray*}{rCl}
  \mathcal{R}(D_1,D_2) \cap \mathcal{C}_{\text{FB}} (P_1,P_2,N) \neq
  \emptyset.
\end{IEEEeqnarray*}
\end{cor}

From the sufficient condition of Corollary \ref{cor:FB-sep-based} and
the necessary condition of Theorem \ref{thm:general-bound} we can now
derive the high-SNR asymptotics of an optimal scheme. To state these
asymptotics, we denote by $(D_1^{\ast},D_2^{\ast})$ an arbitrary
distortion pair resulting from an optimal scheme.

\begin{thm}[High-SNR Distortion] \label{thm:fb-asymptotics} The
  high-SNR asymptotic behavior of $(D_1^{\ast},D_2^{\ast})$ is given
  by
  \begin{IEEEeqnarray*}{rCl}
    \lim_{N \rightarrow 0} \frac{P_1 + P_2 + 2 \sqrt{P_1P_2}}{N}
    D_1^{\ast} D_2^{\ast} & = & \sigma^4 (1-\rho^2),
  \end{IEEEeqnarray*}
  provided that $D_1^{\ast} \leq \sigma^2$ and $D_2^{\ast} \leq
  \sigma^2$, and that
  \begin{IEEEeqnarray}{rCl} \label{eq:fb-lim-Di}
    \lim_{N \rightarrow 0} \frac{N}{P_1 D_1^{\ast}} = 0 & \qquad \text{and} 
    \qquad & \lim_{N \rightarrow 0} \frac{N}{P_2 D_2^{\ast}} = 0.
  \end{IEEEeqnarray}
\end{thm}

\begin{proof}
  See Appendix \ref{appx:prf-high-SNR-FB}.
\end{proof}

\begin{rmk}
  The asymptotics of Theorem \ref{thm:fb-asymptotics} are almost the
  same as those in \cite[Theorem~4.5]{lapidoth-tinguely08-mac-it} for
  the setup without feedback.  The only difference is that in the case
  with feedback the power term $P_1 + P_2 + 2 \rho \sqrt{P_1P_2}$ is
  replaced by $P_1 + P_2 + 2 \sqrt{P_1P_2}$. This stems from the fact
  that with feedback, as $P/N \rightarrow \infty$, the cooperation
  between the transmitters can be full.
\end{rmk}

\begin{rmk}
  Note that under source-channel separation, which achieves the
  high-SNR asymptotics, the cooperation between the transmitters
  takes place only at the channel-coding level. The source-coding is
  performed in a distributed manner.
\end{rmk}



To conclude this section we restate Theorem \ref{thm:fb-asymptotics}
more specifically for the symmetric case. Since there $D_1^{\ast} =
D_2^{\ast} = D^{\ast}(\sigma^2, \rho, P, N)$, condition
\eqref{eq:fb-lim-Di} is implicitly satisfied. Thus,
\begin{cor}\label{cor:fb-sym-asmptotics}
  In the symmetric case
  \begin{IEEEeqnarray*}{rCl}
    \lim_{\frac{P}{N} \rightarrow \infty} \sqrt{\frac{P}{N}}
    D_{\textnormal{FB}}^{\ast}(\sigma^2,\rho,P,N) & = & \sigma^2
    \sqrt{\frac{1-\rho^2}{4}}.
  \end{IEEEeqnarray*}
\end{cor}

\subsection{Uncoded Scheme}\label{subsec:uncoded}

We now revisit the uncoded scheme of \cite[Section
4.3]{lapidoth-tinguely08-mac-it}, which was shown to be optimal for
the setup without feedback whenever the SNR is below a certain
threshold. For our setup with feedback, we show that this scheme is
still optimal whenever the SNR is below the threshold of \cite[Section
4.3]{lapidoth-tinguely08-mac-it}. This result implies that below this
SNR-threshold feedback is useless.\footnote{By the simple structure of
  the uncoded scheme, it follows that feedback is useless not only in
  terms of performance, but also in terms of delay and
  complexity.} Note, however, that feedback is beneficial for the
source-channel separation approach because, even if noisy, it
increases the capacity region of the Gaussian multiple-access channel
\cite{lapidoth-wigger06}.

The uncoded scheme operates as follows. Encoder $i \in \{ 1,2 \}$
produces a time-$k$ channel input $X_{i,k}$ which is a scaled version
of the time-$k$ source output $S_{i,k}$. The scaling is such that the
average power constraint of the channel \eqref{eq:power-constraint} is
satisfied. That is,
\begin{IEEEeqnarray*}{rCl}
  \hspace{40mm} X_{i,k}^{\textnormal{u}} & = &
  \sqrt{\frac{P_i}{\sigma^2}} S_{i,k} \qquad \text{for all } k \in \{
  1,2, \ldots ,n \}.
\end{IEEEeqnarray*}
The decoder reconstructs the source output $S_{i,k}$ by performing the
MMSE estimate of $S_{i,k}$, $i \in \{ 1,2 \}$, $k \in \{ 1,2, \ldots
,n \}$, based on the time-$k$ channel output $Y_k$. That is,
\begin{IEEEeqnarray*}{rCl}
  \hat{S}_{i,k}^{\textnormal{u}} & = & \E{ S_{i,k} | Y_k }.
\end{IEEEeqnarray*}
The expected distortions $(D_1^{\textnormal{u}},D_2^{\textnormal{u}})$
resulting from this uncoded scheme as well as its optimality below a
certain SNR-threshold are stated in the following theorem.

\begin{thm}\label{thm:macfb-uncoded}
  The distortion pairs $(D_1^{\textnormal{u}},D_2^{\textnormal{u}})$
  resulting from the described uncoded scheme are given by
  \begin{IEEEeqnarray*}{rCl}
    D_1^{\textnormal{u}} = \sigma^2 \frac{(1-\rho^2)P_2 + N}{P_1 + P_2 +
      2\rho \sqrt{P_1P_2} + N} & \qquad \qquad & D_2^{\textnormal{u}} =
    \sigma^2 \frac{(1-\rho^2)P_1 + N}{P_1 + P_2 + 2\rho \sqrt{P_1P_2} +
      N}.
  \end{IEEEeqnarray*}
  These distortion pairs $(D_1^{\textnormal{u}},D_2^{\textnormal{u}})$
  are optimal, i.e., lie on the boundary of the distortion region,
  whenever
  \begin{IEEEeqnarray}{rCl}\label{eq:mac-SNRcond-uc-opt}
    P_2 (1-\rho^2)^2 \Big( P_1 + 2\rho \sqrt{P_1P_2} \Big) & \leq &
    N\rho^2 \Big( 2 P_2 (1-\rho^2) + N \Big).
  \end{IEEEeqnarray}
\end{thm}

\begin{proof}
  The expressions for $D_1^{\textnormal{u}}$ and
  $D_2^{\textnormal{u}}$ are derived in \cite[Appendix
  D]{lapidoth-tinguely08-mac-it}. The optimality of the uncoded scheme
  is proven in Appendix \ref{appx:prf-opt-uncoded}. For the particular
  case where $P_1$, $P_2$, $N$ satisfy \eqref{eq:mac-SNRcond-uc-opt}
  with equality, the optimality can also be verified directly from
  Theorem~\ref{thm:general-bound}. To this end, it suffices to notice
  that for $(D_1,D_2) = (D_1^{\textnormal{u}}, D_2^{\textnormal{u}})$,
  the necessary condition of Theorem~\ref{thm:general-bound} is
  satisfied with equality for $\hat{\rho}^{\ast} = \rho$. It thus
  follows that for any $(D'_1,D'_2)$ satisfying $D'_1 \leq
  D_1^{\textnormal{u}}$ and $D'_2 < D_2^{\textnormal{u}}$ or $D'_1 <
  D_1^{\textnormal{u}}$ and $D'_2 \leq D_2^{\textnormal{u}}$ the
  necessary condition of Theorem~\ref{thm:general-bound} is violated
  for every $\hat{\rho} \in [-1,1]$. And hence, the uncoded scheme is
  optimal.
\end{proof}

\begin{cor}
  Source-channel separation is in general suboptimal.
\end{cor}

\begin{proof}
  This can be verified by comparing the achievable distortions given
  in Corollary~\ref{cor:FB-sep-based} with the achievable distortions
  given in Theorem~\ref{thm:macfb-uncoded}. For example, in the
  symmetric case it can be verified that for all $\rho>0$ and $P/N
  \leq \rho/(1-\rho^2)$, the smallest distortions achievable by
  source-channel separation (Corollary~\ref{cor:FB-sep-based}) are
  strictly larger than the distortions resulting from the optimal
  uncoded scheme (Theorem~\ref{thm:macfb-uncoded}).
\end{proof}

\begin{rmk}
  From Theorem \ref{thm:macfb-uncoded} it follows that if $P_1$,
  $P_2$, $N$ satisfy \eqref{eq:mac-SNRcond-uc-opt} with a strict
  inequality, then the necessary condition of Theorem
  \ref{thm:general-bound} is not sufficient. This is due to the
  constraints \eqref{eq:thm-cond2} and \eqref{eq:thm-cond3} which are
  loose at low SNRs and is best seen in the symmetric case. In the
  symmetric case, Theorem \ref{thm:general-bound}
  (cf.~\eqref{eq:thm-cond2} and \eqref{eq:thm-cond3}) yields that for
  $(D,D)$ to be achievable, it is necessary that $D$ satisfy
  \begin{IEEEeqnarray}{rCl}\label{eq:rmk-lb-subopt}
    D & \geq & \sigma^2 (1-\rho^2) \frac{N}{N + P(1-\hat{\rho}^2)},
  \end{IEEEeqnarray}
  i.e., that \eqref{eq:sym-cond2} hold. Since $\hat{\rho} \in [0,1]$,
  the RHS of \eqref{eq:rmk-lb-subopt} is upper bounded by
  ${\sigma^2(1-\rho^2)}$. Thus, for sufficiently low SNRs the constraint
  of \eqref{eq:rmk-lb-subopt} is inactive, and the only active
  constraint is the one of \eqref{eq:sym-cond1}. But, if only
  \eqref{eq:sym-cond1} is active, then $\hat{\rho}^{\ast} = 1$, which
  corresponds to fully cooperating transmitters, and thus, yields a
  loose lower bound on $D^{\ast}(\sigma^2, \rho, P, N)$ at low SNRs.
\end{rmk}


We conclude the section on our main results by restating Theorem
\ref{thm:macfb-uncoded} more specifically for the symmetric case.
\begin{cor}\label{cor:sym-case}
  In the symmetric case
  \begin{IEEEeqnarray}{rCl}\label{eq:cor-symmetric}
    D^{\ast}(\sigma^2,\rho,P,N) & = & \sigma^2
    \frac{P(1-\rho^2) + N}{2P(1+\rho) + N}, \qquad \quad \frac{P}{N} \leq
    \frac{\rho}{1-\rho^2}.
  \end{IEEEeqnarray}
\end{cor}

\section{Summary}

We studied the power-versus-distortion trade-off for the transmission
of a memoryless bivariate Gaussian source over a two-to-one
average-power limited Gaussian multiple-access channel with perfect
causal feedback. In this problem, each of two separate transmitters observes
a different component of a memoryless bivariate Gaussian source
as well as the feedback from the channel output of the previous
time-instants. Based on the observed source sequence and the feedback,
each transmitter then describes its source component to the common
receiver via an average-power constrained Gaussian multiple-access
channel. From the resulting channel output, the receiver wishes to
reconstruct both source components with the least possible expected
squared-error distortion. Our interest was in the set of distortion
pairs that can be achieved by the receiver on the two source
components.
%
%
%
%
Our main results were:\\

\begin{itemize}
\item A necessary condition (Theorem~\ref{thm:general-bound}) for the
  achievability of a distortion pair $(D_1,D_2)$.\\
\item The precise high-SNR asymptotic behaviour (Theorem
  \ref{thm:fb-asymptotics}) of optimal transmission schemes, which in
  the symmetric case (Corollary \ref{cor:fb-sym-asmptotics}) is given
  by
  \begin{IEEEeqnarray*}{rCl}
    \lim_{P/N \rightarrow \infty} \sqrt{\frac{P}{N}}
    D^{\ast}(\sigma^2,\rho,P,N) & = & \sigma^2
    \sqrt{\frac{1-\rho^2}{4}},
  \end{IEEEeqnarray*}
  and which is shown to be achievable by source-channel separation.\\
\item The optimality, for all SNRs below a certain threshold, of an
  uncoded transmission scheme, which ignores the feedback
  (Theorem~\ref{thm:macfb-uncoded}). In the symmetric case, this
  optimality result (Corollary~\ref{cor:sym-case}) is given by
  \begin{IEEEeqnarray*}{rCl}
    D^{\ast}(\sigma^2,\rho,P,N) = \sigma^2
    \frac{P(1-\rho^2)+N}{2P(1+\rho)+N}, & \qquad \quad
    \frac{P}{N} \leq \frac{\rho}{1-\rho^2}.
  \end{IEEEeqnarray*}
\end{itemize}

\appendix

\section{Proof of Theorem
  \ref{thm:general-bound}} \label{appx:prf-thm-lbFB}

In Theorem \ref{thm:general-bound} we have given a necessary condition
for the achievability of a distortion pair $(D_1,D_2)$ for the
multiple-access problem with feedback. The proof of this necessary
condition uses the following two lemmas.
\begin{lm}\label{lm:macfb-ub-RD}
  For our multiple-access setup with feedback, let $\{ X_{1,k} \}$,
  $\{ X_{2,k} \}$ and $\{ Y_k \}$ be the channel inputs and channel
  outputs of a coding scheme achieving some distortion pair
  $(D_1,D_2)$. Then, for every $\delta > 0$ there exists an
  $n_0(\delta) > 0$ such that for all $n > n_0(\delta)$
  \begin{IEEEeqnarray}{rCl}
    n R_{S_1,S_2}(D_1+\delta,D_2+\delta) & \leq & \sum_{k=1}^n
    I(X_{1,k},X_{2,k};Y_k), \label{eq:prf-sum-rate} \\
    n R_{S_1|S_2}(D_1+\delta) & \leq & \sum_{k=1}^n I(X_{1,k};Y_k|X_{2,k}),
    \label{eq:prf-side-info1} \\
    n R_{S_2|S_1}(D_2+\delta) & \leq & \sum_{k=1}^n I(X_{2,k};Y_k|X_{1,k}).
    \label{eq:prf-side-info2}
  \end{IEEEeqnarray}
\end{lm}

\begin{proof}
  The proofs of \eqref{eq:prf-sum-rate} -- \eqref{eq:prf-side-info2}
  follow along the lines of the proof for the univariate analog (see
  e.g.~\cite[page 15]{gastpar-thesis}). The main ingredients in those
  derivations are the convexity of the rate-distortion functions and
  the data-processing inequality. We start with the proof of
  \eqref{eq:prf-sum-rate}. By the definition of an achievable
  distortion pair $(D_1,D_2)$ (Definition \ref{def:achv-dist}) and by
  the monotonicity of $R_{S_1,S_2}(\Delta_1,\Delta_2)$ in
  $(\Delta_1,\Delta_2)$, we have that for every $\delta > 0$ there
  exists an $n_0(\delta) > 0$ such that for every $n > n_0(\delta)$
  \begin{IEEEeqnarray}{rCl}
    n R_{S_1,S_2}(D_1+\delta,D_2+\delta) & \leq & n R_{S_1,S_2} \left( \frac{1}{n}
      \sum_{k=1}^n \E{(S_{1,k} - \hat{S}_{1,k})^2}, \frac{1}{n}
      \sum_{k=1}^n \E{(S_{2,k} - \hat{S}_{2,k})^2} \right) \nonumber\\
    & \stackrel{a)}{\leq} & n \sum_{k=1}^n \frac{1}{n} R_{S_1,S_2}
    \Big( \underbrace{\E{(S_{1,k} - \hat{S}_{1,k})^2}}_{d_{1,k}},
    \underbrace{\E{(S_{2,k} - \hat{S}_{2,k})^2}}_{d_{2,k}} \Big) \nonumber\\
    & = & \sum_{k=1}^n \min_{\substack{P_{T_1,T_2|S_1,S_2}:\\
        \E{(S_1-T_1)^2} \leq d_{1,k}\\ \E{(S_2-T_2)^2} \leq
        d_{2,k}}} I(S_1,S_2;T_1,T_2) \nonumber\\
    & \leq & \sum_{k=1}^n I(S_{1,k},S_{2,k} ; \hat{S}_{1,k},
    \hat{S}_{2,k}) \nonumber\\
    & = & \sum_{k=1}^n h(S_{1,k},S_{2,k}) - \sum_{k=1}^n
    h(S_{1,k},S_{2,k} | \hat{S}_{1,k},\hat{S}_{2,k}) \nonumber\\
    & \leq & \sum_{k=1}^n h(S_{1,k},S_{2,k}) - \sum_{k=1}^n
    h(S_{1,k},S_{2,k} | \hat{\bf S}_1,\hat{\bf S}_2, S_{1,1}^{k-1},
    S_{2,1}^{k-1}) \nonumber\\
    & = & h({\bf S}_1,{\bf S}_2) - h( {\bf S}_1, {\bf S}_2 | \hat{\bf
      S}_1,\hat{\bf S}_2) \nonumber\\
    & = & I({\bf S}_1, {\bf S}_2 ; \hat{\bf S}_1,\hat{\bf
      S}_2) \nonumber\\
    & \stackrel{b)}{\leq} & I({\bf S}_1, {\bf S}_2 ; {\bf
      Y}), \label{eq:macfb-ub-RD1D2-1} 
  \end{IEEEeqnarray}
  where in step $a)$ we have used of the convexity of
  $R_{S_1,S_2}(D_1,D_2)$, and in step $b)$ we have used the
  data-processing inequality. The RHS of \eqref{eq:macfb-ub-RD1D2-1} can be
  further bounded as follows
  \begin{IEEEeqnarray}{rCl}
    I({\bf S}_1, {\bf S}_2 ; {\bf Y}) & = & h({\bf Y}) - h({\bf Y} |
    {\bf S}_1, {\bf S}_2) \nonumber\\
    & = & h({\bf Y}) - \sum_{k=1}^n h(Y_k | {\bf S}_1, {\bf S}_2,
    Y_1^{k-1}) \nonumber\\
    & \leq & h({\bf Y}) - \sum_{k=1}^n h(Y_k | {\bf S}_1, {\bf S}_2,
    Y_1^{k-1}, X_{1,k}, X_{2,k}) \nonumber\\
    & \stackrel{a)}{=} & h({\bf Y}) - \sum_{k=1}^n h(Y_k | X_{1,k},
    X_{2,k}) \nonumber\\ 
    & \leq & \sum_{k=1}^n h(Y_k) - \sum_{k=1}^n h(Y_k | X_{1,k}, X_{2,k})
    \nonumber\\ 
    & = & \sum_{k=1}^n I(X_{1,k}, X_{2,k};Y_k), \label{eq:macfb-ub-IS1S2S1hS2h}
  \end{IEEEeqnarray}
  where inequality $a)$ follows because given the channel inputs
  $X_{1,k}$, $X_{2,k}$, the channel output $Y_k$ is independent of
  $({\bf S}_1, {\bf S}_2, Y_1^{k-1})$. Inequalities
  \eqref{eq:macfb-ub-RD1D2-1} and \eqref{eq:macfb-ub-IS1S2S1hS2h}
  combine to prove \eqref{eq:prf-sum-rate}.

  The derivations for \eqref{eq:prf-side-info1} and
  \eqref{eq:prf-side-info2} are similar to the one for
  \eqref{eq:prf-sum-rate}. Since there is a symmetry between the
  derivation of \eqref{eq:prf-side-info1} and the derivation of
  \eqref{eq:prf-side-info2}, we only give the derivation of
  \eqref{eq:prf-side-info1}. By the definition of an achievable
  distortion pair $(D_1,D_2)$ and by the monotonicity of
  $R_{S_1|S_2}(\Delta_1)$ in $\Delta_1$, we have that for every
  $\delta > 0$ there exists an $n_0(\delta) > 0$ such that for every
  $n > n_0(\delta)$ we have
  \begin{IEEEeqnarray}{rCl}
    n R_{S_1|S_2}(D_1+\delta) & \leq & n R_{S_1|S_2} \left( \frac{1}{n}
      \sum_{k=1}^n \E{(S_{1,k} - \hat{S}_{1,k})^2} \right) \nonumber\\
    & \stackrel{a)}{\leq} & n \sum_{k=1}^n \frac{1}{n} R_{S_1|S_2} \Big(
    \underbrace{\E{(S_{1,k} - \hat{S}_{1,k})^2}}_{d_{1,k}} \Big) \nonumber\\
    & = & \sum_{k=1}^n \min_{\substack{P_{T_k|S_{1,k},S_{2,k}}:\\
        \E{(S_{1,k}-T_k)^2} \leq d_{1,k} }} I(S_{1,k};T_k|S_{2,k}) \nonumber\\
    & \leq & \sum_{k=1}^n I(S_{1,k} ; \hat{S}_{1,k} | S_{2,k}) \nonumber\\
    & = & \sum_{k=1}^n h(S_{1,k}|S_{2,k}) - \sum_{k=1}^n
    h(S_{1,k} | \hat{S}_{1,k},S_{2,k}) \nonumber\\
    & = & \sum_{k=1}^n h(S_{1,k}|S_{1,1}^{k-1},{\bf S}_2) - \sum_{k=1}^n
    h(S_{1,k} | \hat{S}_{1,k},S_{2,k}) \nonumber\\
    & \leq & \sum_{k=1}^n h(S_{1,k}|S_{1,1}^{k-1},{\bf S}_2) -
    \sum_{k=1}^n
    h(S_{1,k} | \hat{\bf S}_1, {\bf S}_2, S_{1,1}^{k-1}) \nonumber\\
    & = & \sum_{k=1}^n I(S_{1,k} ; \hat{\bf S}_1| {\bf S}_2,
    S_{1,1}^{k-1}) \nonumber\\
    & = & I({\bf S}_1 ; \hat{\bf S}_1 | {\bf S}_2) \nonumber\\
    & \stackrel{b)}{\leq} & I({\bf S}_1, {\bf Y} | {\bf
      S}_2), \label{eq:macfb-ub-RD1-1}
  \end{IEEEeqnarray}
  where step $a)$ follows by the convexity of $R_{S_1|S_2}(D_1)$ and
  step $b)$ follows by the data-processing in equality, i.e.
  \begin{IEEEeqnarray*}{rCl}
    I({\bf S}_1;{\bf Y},\hat{\bf S}_1|{\bf S}_2) & = & I({\bf S}_1;\hat{\bf
      S}_1|{\bf S}_2) - \underbrace{I({\bf S}_1;{\bf Y}|\hat{\bf S}_1, {\bf
        S}_2)}_{\geq 0}\\  
    & = & I({\bf S}_1;{\bf Y}|{\bf S}_2) - \underbrace{I({\bf S}_1;\hat{\bf
        S}_1 | {\bf Y}, {\bf S}_2)}_{=0}.
  \end{IEEEeqnarray*}
  The RHS of \eqref{eq:macfb-ub-RD1-1} can be further bounded as follows
  \begin{IEEEeqnarray}{rCl}
    I({\bf S}_1 ; {\bf Y} | {\bf S}_2) & = & h({\bf Y} | {\bf
      S}_2) - h({\bf Y} | {\bf S}_1, {\bf S}_2) \nonumber\\
    & = & \sum_{k=1}^n h(Y_k | Y_1^{k-1}, {\bf S}_2) - \sum_{k=1}^n
    h(Y_k | {\bf S}_1, {\bf S}_2, Y_1^{k-1}) \nonumber\\
    & = & \sum_{k=1}^n h(Y_k | Y_1^{k-1}, {\bf S}_2, X_{2,k}) -
    \sum_{k=1}^n h(Y_k | {\bf S}_1, {\bf S}_2, Y_1^{k-1}, X_{1,k}, X_{2,k}) \nonumber\\
    & \stackrel{a)}{\leq} & \sum_{k=1}^n h(Y_k | X_{2,k}) - \sum_{k=1}^n h(Y_k |
    X_{1,k}, X_{2,k}) \nonumber\\
    & = & \sum_{k=1}^n I(X_{1,k} ; Y_k | X_{2,k}), \label{eq:IS1|S2-IX1|X2}
  \end{IEEEeqnarray}
  where $a)$ follows because given the channel inputs $X_{1,k}$,
  $X_{2,k}$, the channel output $Y_k$ is independent of $({\bf S}_1,
  {\bf S}_2, Y_1^{k-1})$. Inequalities \eqref{eq:macfb-ub-RD1-1} and
  \eqref{eq:IS1|S2-IX1|X2} combine to prove \eqref{eq:prf-side-info1}.
\end{proof}

\begin{lm}\label{lemma:MAC-rate}
  Let $\{ X_{1,k} \}$ and $\{ X_{2,k} \}$ be zero-mean sequences
  satisfying $\sum_{i=1}^n \E{X_{i,k}^2} \leq n P_i$,
  $i~\in~\{1,2\}$. Let $Y_k = X_{1,k} + X_{2,k} + Z_k$, where
  $\{Z_k\}$ are IID zero-mean variance-$N$ Gaussian, and where for
  every $k$, $Z_k$ is independent of $(X_{1,k}, X_{2,k})$. Let
  $\hat{\rho}_n \in [0,1]$ be given by
  \begin{equation}\label{eq:tilde-rho}
    \hat{\rho}_n \triangleq \frac{\left| \frac{1}{n} \sum_{k=1}^n
        \E{X_{1,k}X_{2,k}} \right|}{\sqrt{\left( \frac{1}{n} \sum_{k=1}^n
          \E{X_{1,k}^2} \right) \left( \frac{1}{n} \sum_{k=1}^n
          \E{X_{2,k}^2} \right)}}.
  \end{equation}
  Then
  \begin{align}
    \sum_{k=1}^n I(X_{1,k},X_{2,k} ; Y_k) &\leq \frac{n}{2} \log_2
    \left( 1 + \frac{P_1 + P_2 + 2 \hat{\rho}_n \sqrt{P_1P_2}}{N}
    \right), \label{eq:MAC-sum-rate}\\ 
    \sum_{k=1}^n I(X_{1,k} ; Y_k|X_{2,k}) &\leq \frac{n}{2} \log_2
    \left( 1 + \frac{P_1(1-\hat{\rho}_n^2)}{N}
    \right),\label{eq:MAC-rate1}\\ 
    \sum_{k=1}^n I(X_{2,k} ; Y_k|X_{1,k}) &\leq \frac{n}{2} \log_2
    \left( 1 + \frac{P_2(1-\hat{\rho}_n^2)}{N}
    \right).\label{eq:MAC-rate2}
  \end{align}
\end{lm}

\begin{proof}
  See \cite[pp.~627]{ozarow85}.
\end{proof}

\begin{proof}[Proof of Theorem \ref{thm:general-bound}]
  The proof now follows by jointly bounding the expressions on the RHS
  of \eqref{eq:prf-sum-rate}, \eqref{eq:prf-side-info1}, and
  \eqref{eq:prf-side-info2} by means of Lemma \ref{lemma:MAC-rate},
  and using that for $n \rightarrow \infty$ Lemma \ref{lm:macfb-ub-RD}
  holds for every $\delta > 0$.
\end{proof}

\section{Proof of Theorem \ref{thm:fb-asymptotics}} \label{appx:prf-high-SNR-FB}

For $\rho = 1$ the result follows by noting that the multiple-access
problem reduces to a point-to-point problem where $D_1^{\ast} =
D_2^{\ast}$. Hence, we shall now assume
\begin{IEEEeqnarray}{rCl}\label{eq:macfb-hsnr-rho<1}
\rho < 1.
\end{IEEEeqnarray}
The result can then be obtained from the necessary condition for the
achievability of a distortion pair $(D_1,D_2)$ in Theorem
\ref{thm:general-bound} and from the sufficient conditions for the
achievability of a distortion pair $(D_1,D_2)$ that follow from
source-channel separation in Corollary \ref{cor:FB-sep-based}.

By Corollary~\ref{cor:FB-sep-based} it follows that a distortion pair
$(\bar{D}_1,\bar{D}_2)$ is achievable if $\bar{D}_1 \leq \sigma^2$,
$\bar{D}_2 \leq \sigma^2$ and 
\begin{IEEEeqnarray}{rCl}
  \bar{D}_1 & \geq & \sigma^2 2^{-2R_1} (1-\rho^2) + \sigma^2 \rho^2
  2^{-2(R_1+R_2)} \label{eq:oohama-D1}\\[2mm]
  \bar{D}_2 & \geq & \sigma^2 2^{-2R_2} (1-\rho^2) + \sigma^2 \rho^2
  2^{-2(R_1+R_2)} \label{eq:oohama-D2}\\[2mm]
  \bar{D}_1 \bar{D}_2 & = & \sigma^4 2^{-2 (R_1+R_2)} (1-\rho^2) + \sigma^4 \rho^2
  2^{-4(R_1 + R_2)}, \label{eq:oohama-D1D2} 
\end{IEEEeqnarray}
where the rate-pair $(R_1,R_2)$ satisfies for some $\bar{\rho} \in [0,1]$
\begin{IEEEeqnarray}{rCl}
  R_1 & \leq & \frac{1}{2} \log_2 \left( \frac{P_1
      (1-\bar{\rho}^2)}{N} \right) \label{eq:ozarow-R1}\\[2mm]
  R_2 & \leq & \frac{1}{2} \log_2 \left( \frac{P_2
      (1-\bar{\rho}^2)}{N} \right) \label{eq:ozarow-R2} \\[2mm]
  R_1 + R_2 & \leq & \frac{1}{2} \log_2 \left( \frac{P_1 + P_2 + 2
      \bar{\rho} \sqrt{P_1P_2}}{N} \right). \label{eq:ozarow-R1R2}
\end{IEEEeqnarray}
If we restrict ourselves to distortion pairs $(\bar{D}_1,\bar{D}_2)$ satisfying
\begin{IEEEeqnarray}{rCl}\label{eq:macfb-D1D2-asymptotic}
  \lim_{N \rightarrow 0} \frac{N}{P_1 \bar{D}_1} = 0 & \qquad \text{and} \qquad &
  \lim_{N \rightarrow 0} \frac{N}{P_2 \bar{D}_2} = 0,
\end{IEEEeqnarray}
and to $\rho$ satisfying \eqref{eq:macfb-hsnr-rho<1}, then for
sufficiently small $N>0$ the constraints \eqref{eq:oohama-D1} and
\eqref{eq:oohama-D2} become redundant. Consequently, for $N$
sufficiently small, any distortion pair $(\bar{D}_1,\bar{D}_2)$
satisfying \eqref{eq:macfb-D1D2-asymptotic} and
\eqref{eq:oohama-D1D2}, where $(R_1,R_2)$ satisfies
\eqref{eq:ozarow-R1}--\eqref{eq:ozarow-R1R2} for some $\bar{\rho} \in
[0,1]$, is achievable. And because for any fixed $\bar{\rho} \in
[0,1)$ as $N \rightarrow 0$ the Constraints \eqref{eq:ozarow-R1} and
\eqref{eq:ozarow-R2} become redundant, it follows that any distortion
pair satisfying \eqref{eq:macfb-D1D2-asymptotic} and
\begin{IEEEeqnarray}{rCl}\label{eq:macfb-D1D2-hsnr1}
  \lim_{N \rightarrow 0} \frac{P_1+P_2+2 \bar{\rho} \sqrt{P_1P_2}}{N}
  \bar{D}_1 \bar{D}_2 & = & \sigma^4 (1-\rho^2),
\end{IEEEeqnarray}
for some $\bar{\rho} \in [0,1)$, is achievable. Since $\bar{\rho}$ can
be chosen arbitrarily close to $1$, a simple calculus argument shows
that
\begin{IEEEeqnarray}{rCl}\label{eq:macfb-D1D2-hsnr2}
  \lim_{N \rightarrow 0} \frac{P_1+P_2+2 \sqrt{P_1P_2}}{N}
  \bar{D}_1 \bar{D}_2 & = & \sigma^4 (1-\rho^2),
\end{IEEEeqnarray}
is achievable.



Next, let $\big( D_1^{\ast} (\sigma^2,\rho,P_1,P_2,N), D_2^{\ast}
(\sigma^2,\rho,P_1,P_2,N) \big)$ be a distortion pair resulting from
an arbitrary optimal scheme for the corresponding SNR, and let
$(D_1^{\ast},D_2^{\ast})$ be the corresponding shorthand notation for
this distortion pair. By Theorem \ref{thm:general-bound} we have that
\begin{IEEEeqnarray}{rCl}\label{eq:macfb-hsnr1}
  R_{S_1,S_2}(D_1,D_2) & \leq & \frac{1}{2} \log_2 \left( 1 + \frac{P_1
      + P_2 + 2 \sqrt{P_1P_2}}{N} \right).
\end{IEEEeqnarray}
If $(D_1^{\ast}, D_2^{\ast})$ satisfies
\begin{IEEEeqnarray}{rCl}\label{eq:macfb-D1D2^*-asymptotic}
  \lim_{N \rightarrow 0} \frac{N}{P_1 D_1^{\ast}} = 0 & \qquad \text{and} \qquad &
  \lim_{N \rightarrow 0} \frac{N}{P_2 D_2^{\ast}} = 0,
\end{IEEEeqnarray}
then for $N$ sufficiently small
\begin{IEEEeqnarray}{rCl}\label{eq:macfb-hsnr-RS1S2-D2}
  R_{S_1,S_2}(D_1^{\ast},D_2^{\ast}) & = & \frac{1}{2} \log_2^+ \left(
    \frac{\sigma^4(1-\rho^2)}{D_1^{\ast}D_2^{\ast}} \right),
\end{IEEEeqnarray}
by Theorem \ref{thm:rd1d2-main} and because $(D_1^{\ast},D_2^{\ast})
\in \mathscr{D}_2$. From \eqref{eq:macfb-hsnr1} and
\eqref{eq:macfb-hsnr-RS1S2-D2} we thus get that if
$(D_1^{\ast},D_2^{\ast})$ satisfies \eqref{eq:macfb-D1D2^*-asymptotic},
then
\begin{IEEEeqnarray}{rCl}\label{eq:macfb-hsnr-opt2}
  \lim_{N \rightarrow 0} \frac{P_1 + P_2 + 2 \sqrt{P_1P_2}}{N}
  D_1^{\ast} D_2^{\ast} & \geq & \sigma^4 (1-\rho^2).
\end{IEEEeqnarray}
Combining \eqref{eq:macfb-D1D2-hsnr2} with \eqref{eq:macfb-hsnr-opt2}
yields Theorem \ref{thm:fb-asymptotics}. \hfill $\Box$

\section{Proof of Theorem
  \ref{thm:macfb-uncoded}} \label{appx:prf-opt-uncoded}

Theorem \ref{thm:macfb-uncoded} states that for the multiple-access
problem with feedback, if $P_1$, $P_2$, $N$ satisfy
\eqref{eq:mac-SNRcond-uc-opt}, then the uncoded scheme is optimal,
i.e.~no pair $(D_1,D_2)$ satisfying $D_1 \leq D_1^{\textnormal{u}}$
and $D_2 < D_2^{\textnormal{u}}$ or satisfying $D_1 <
D_1^{\textnormal{u}}$ and $D_2 \leq D_2^{\textnormal{u}}$ is
achievable. For $P_1$, $P_2$, $N$ satisfying
\eqref{eq:mac-SNRcond-uc-opt} with equality this was proven right
after Theorem \ref{thm:macfb-uncoded}. Thus, here we restrict
ourselves to $P_1$, $P_2$, $N$ satisfying
\eqref{eq:mac-SNRcond-uc-opt} with strict inequality.

We now show the inachievability of every $(D_1,D_2)$ satisfying $D_1 <
D_1^{\textnormal{u}}$ and $D_2 \leq D_2^{\textnormal{u}}$. The
inachievability of every $(D_1,D_2)$ satisfying $D_1 \leq
D_1^{\textnormal{u}}$ and $D_2 < D_2^{\textnormal{u}}$ follows by
similar arguments and is therefore omitted. The main step in our proof
follows by contradiction. More precisely, we show that a contradiction
arises from the following assumption.
\begin{assumption}[Leading to a contradiction]
  \label{as:false}
  For $P_1$, $P_2$, $N$ satisfying \eqref{eq:mac-SNRcond-uc-opt} with
  strict inequality, there exist encoding rules $\{f_{i,k}^{(n)}\}$
  satisfying the average power constraints
  \eqref{eq:power-constraint}, which, when combined with the optimal
  conditional expectation reconstructors
  \begin{IEEEeqnarray}{rCl}\label{eq:macfb-hatSi}
    \hat{\bf S}_i & = & \E{{\bf S}_i | {\bf Y}}, \qquad i \in \{ 1,2 \},
  \end{IEEEeqnarray}
  result in
  \begin{IEEEeqnarray}{rCl}\label{eq:macfb-def-Di*}
    \varlimsup_{n \rightarrow \infty} \frac{1}{n} \sum_{k=1}^n
    \E{(S_{i,k} - \hat{S}_{i,k})^2} & \triangleq & D_i^{\ast} \qquad
    \qquad i \in \{ 1,2 \},
  \end{IEEEeqnarray}
  such that
  \begin{IEEEeqnarray}{rCl}\label{eq:macfb-ass-contr}
    (D_1^{\ast},D_2^{\ast}) \in \textnormal{int}(\mathscr{D}_3), \qquad \quad
    D_1^{\ast} < D_1^{\textnormal{u}} & \quad \qquad \text{and}
    \quad \qquad &
    D_2^{\ast} = D_2^{\textnormal{u}},
  \end{IEEEeqnarray}
  where we have denoted by $\textnormal{int}(\mathscr{D}_3)$ the
  interior of $\mathscr{D}_3$.
\end{assumption}
Once a contradiction from Assumption~\ref{as:false} is established, it
will follow that Assumption~\ref{as:false} is false and the proof of
Theorem~\ref{thm:macfb-uncoded} will follow in Section
\ref{sec:macfb-uc-prf-conclude}.

Assume that Assumption~\ref{as:false} is true. Let $\{f_{i,k}^{(n)}\}$
be a sequence of encoding functions, with resulting channel inputs $\{
X_{1,k}, X_{2,k} \}$ and resulting channel outputs $\{ Y_k \}$, which,
when combined with the optimal conditional expectation reconstructors
$\hat{\bf S}_1 = \E{ {\bf S}_1 \big| {\bf Y}}$ and $\hat{\bf S}_2 =
\E{ {\bf S}_2 \big| {\bf Y}}$ result in distortions
$(D_1^{\ast},D_2^{\ast})$ as defined in \eqref{eq:macfb-def-Di*} and
satisfying \eqref{eq:macfb-ass-contr}. The contradiction based on
Assumption~\ref{as:false} will be obtained by deriving contradictory
lower and upper bounds for the expected squared-error that
Transmitter~2 can achieve at the end of the transmission on the
sequence ${\bf W} \triangleq {\bf S}_1 - \rho {\bf S}_2$. To this end,
let $\varphi^{(n)} ({\bf S}_2,{\bf Y})$ be some estimator of ${\bf W}$
from $({\bf S}_2,{\bf Y})$ and let $D_W(\varphi^{(n)})$ be the mean
squared-error associated with it:
\begin{IEEEeqnarray*}{rCl}
  D_W(\varphi^{(n)}) & \triangleq & \frac{1}{n} \E{\| {\bf W} -
    \varphi^{(n)}({\bf S}_2,{\bf Y}) \|^2}.
\end{IEEEeqnarray*}
Based on Assumption \ref{as:false}, we now derive a lower bound on
$D_W(\varphi^{(n)})$.

\subsection{``Lower Bound'' on $D_W(\varphi^{(n)})$}\label{subsec:prf-macfb-lb-uc}
In this section we show that
\begin{IEEEeqnarray}{rCl}\label{eq:lb-Dw}
  \text{Assumption \ref{as:false} } & \hspace{2mm} \Rightarrow \hspace{2mm} &
  \left( \varliminf_{n \rightarrow \infty} D_W(\varphi^{(n)}) > \sigma^2
  (1-\rho^2) \frac{N}{N + P_1(1-\rho^2)} \qquad \forall
  \varphi^{(n)} \right). \hspace{8mm}
\end{IEEEeqnarray}
The idea in showing \eqref{eq:lb-Dw} is to exploit the fact that the
sequence ${\bf W}$ is independent of ${\bf S}_2$, and that therefore
the only information that Transmitter~2 receives about ${\bf W}$ is
via the feedback signal ${\bf Y}$. Roughly speaking, we then show that
if ${\bf Y}$ allows for ``good'' estimates of ${\bf S}_1$ and ${\bf
  S}_2$, i.e.~if $D^{\ast}_1 < D^{\textnormal{u}}_1$ and $D^{\ast}_2 =
D^{\textnormal{u}}_2$, then ${\bf Y}$ can only contain ``little''
information about ${\bf W}$, and hence Transmitter~2 can only make a
coarse estimate of ${\bf W}$. The main element in showing
\eqref{eq:lb-Dw} is given by the following lemma.
\begin{lm}\label{lemma:mut-info}
  Let $\hat{\rho}_n$ be as defined in \eqref{eq:tilde-rho}. Then
  \begin{IEEEeqnarray*}{rCl}
    I({\bf S}_1;{\bf Y} | {\bf S}_2) & \leq & \frac{n}{2} \log_2 \left( 1 +
      \frac{P_1(1-\hat{\rho}_n^2)}{N} \right)
  \end{IEEEeqnarray*}
  and
  \begin{IEEEeqnarray*}{rCl}
    \textnormal{Assumption \ref{as:false} } & \quad \Rightarrow \quad
    & \varliminf_{n \rightarrow \infty} \hat{\rho}_n > \rho.
  \end{IEEEeqnarray*}
\end{lm}

\begin{proof}
  Combining \eqref{eq:IS1|S2-IX1|X2} with Lemma~\ref{lemma:MAC-rate}
  Inequality~\eqref{eq:MAC-rate1} gives
  \begin{IEEEeqnarray*}{rCl}
    I({\bf S}_1;{\bf Y}|{\bf S}_2) & \leq & \frac{n}{2} \log_2 \left( 1
      + \frac{P_1(1-\hat{\rho}_n^2)}{N} \right),
  \end{IEEEeqnarray*}
  with $\hat{\rho}_n$ as defined in \eqref{eq:tilde-rho}. It now
  remains to show that Assumption \ref{as:false} implies that
  $\varliminf_{n \rightarrow \infty} \hat{\rho}_n > \rho$. To this
  end, we recall that from \cite[Proof of Theorem
  4.1]{lapidoth-tinguely08-mac-it} we have that if $P_1$, $P_2$, $N$
  satisfy \eqref{eq:mac-SNRcond-uc-opt}, then the corresponding
  $(D_1^{\textnormal{u}}, D_2^{\textnormal{u}})$ satisfies
  \cite[Condition (14) of Theorem 4.1]{lapidoth-tinguely08-mac-it}
  with equality, i.e.,
  \begin{IEEEeqnarray}{rCl}\label{eq:macfb-D1uD2u=C}
    R_{S_1,S_2}(D^{\textnormal{u}}_1,D^{\textnormal{u}}_2) & = & \frac{1}{2}
    \log_2 \left( 1 + \frac{P_1 +P_2 +2\rho \sqrt{P_1P_2}}{N} \right).
  \end{IEEEeqnarray}
  Next, we notice that since Assumption \ref{as:false} guarantees that
  $(D_1^{\ast},D_2^{\ast})$ is achievable, it follows from Lemma
  \ref{lm:macfb-ub-RD} that for every $\delta > 0$ there exists an
  $n'(\delta) > 0$ such that for all $n > n'(\delta)$ we have
  \begin{IEEEeqnarray}{rCl}
    n R_{S_1,S_2}(D^{\ast}_1+\delta,D^{\ast}_2+\delta) & \stackrel{a)}{\leq} &
    \sum_{k=1}^n I(X_{1,k},X_{2,k} ;
    Y_k) \qquad \qquad \nonumber \\
    & \stackrel{b)}{\leq} & \frac{n}{2} \log_2 \left( 1 + \frac{P_1 +
        P_2 + 2\hat{\rho}_n \sqrt{P_1P_2}}{N}\right),\label{eq:RD-sum-rate}
  \end{IEEEeqnarray}
  where $a)$ follows from \eqref{eq:prf-sum-rate} in
  Lemma~\ref{lm:macfb-ub-RD}, and $b)$ follows from Lemma
  \ref{lemma:MAC-rate}. Taking the $\liminf$ of \eqref{eq:RD-sum-rate}
  yields that for every $\delta > 0$
  \begin{IEEEeqnarray*}{rCl}
    R_{S_1,S_2}(D^{\ast}_1+\delta,D^{\ast}_2+\delta) & \leq &
    \frac{1}{2} \log_2 \left( 1 + \frac{P_1 + P_2 +
        2\hat{\rho}^{\ast} \sqrt{P_1P_2}}{N}\right),
  \end{IEEEeqnarray*}
  where $\hat{\rho}^{\ast} = \varliminf_{n \rightarrow \infty}
  \hat{\rho}_n$. And since $R_{S_1,S_2}(D_1,D_2)$ is continuous in
  $(D_1,D_2)$ it follows, upon letting $\delta$ tend to zero, that
  \begin{IEEEeqnarray}{rCl}\label{eq:macfb-uc-rhohat*}
    R_{S_1,S_2}(D^{\ast}_1,D^{\ast}_2) & \leq & \frac{1}{2} \log_2
    \left( 1 + \frac{P_1 + P_2 + 2\hat{\rho}^{\ast}
        \sqrt{P_1P_2}}{N}\right).
  \end{IEEEeqnarray}
  By Assumption \ref{as:false} and by the strict monotonicity of
  $R_{S_1,S_2}(D_1,D_2)$ as a function of $D_1$ in
  $\textnormal{int}(\mathscr{D}_3)$, it follows from the hypothesis
  $D_1^{\ast} < D_1^{\textnormal{u}}$ and $D_1^{\ast} =
  D_1^{\textnormal{u}}$ that
  \begin{IEEEeqnarray}{rCl}\label{eq:macfb-RD1uD2u-RD1*D2*}
    R_{S_1,S_2}(D^{\textnormal{u}}_1,D^{\textnormal{u}}_2) & < &
    R_{S_1,S_2}(D^{\ast}_1,D^{\ast}_2).
  \end{IEEEeqnarray}
  Combining \eqref{eq:macfb-RD1uD2u-RD1*D2*} with
  \eqref{eq:macfb-uc-rhohat*} and \eqref{eq:macfb-D1uD2u=C} gives
  $\varliminf_{n \rightarrow \infty} \hat{\rho}_n > \rho$.
\end{proof}

We next prove that
\begin{IEEEeqnarray}{rCl}\label{eq:mut-inf-bd}
D_W(\varphi^{(n)}) & \geq & \sigma^2 (1-\rho^2) 2^{-\frac{2}{n} I({\bf
    S}_1;{\bf Y}|{\bf S}_2)}.
\end{IEEEeqnarray}
To derive \eqref{eq:mut-inf-bd}, denote by $R_W(D)$ the
rate-distortion function for a source of the law of ${\bf W}$. We then
have \vspace{-2mm}
\begin{IEEEeqnarray}{rCl}
n R_W(D_W(\varphi^{(n)})) & \stackrel{a)}{\leq} &
I({\bf W};\varphi^{(n)}({\bf S}_2,{\bf Y})) \nonumber \\ 
& \stackrel{b)}{\leq} & I({\bf W};{\bf Y},{\bf S}_2) \nonumber \\
& = & I({\bf S}_1-\rho {\bf S}_2;{\bf Y},{\bf S}_2) \nonumber \\
& = & h({\bf S}_1-\rho {\bf S}_2) - h({\bf S}_1-\rho {\bf S}_2 | {\bf
  Y},{\bf S}_2) \nonumber \\ 
& \stackrel{c)}{=} & h({\bf S}_1-\rho {\bf S}_2|{\bf S}_2) - h({\bf
  S}_1-\rho {\bf S}_2 | {\bf Y},{\bf S}_2)\nonumber  \\
& = & h({\bf S}_1|{\bf S}_2) - h({\bf S}_1|{\bf Y},{\bf S}_2) \nonumber \\
& = & I({\bf S}_1;{\bf Y}|{\bf S}_2), \label{eq:mut-inf-Dw}
\end{IEEEeqnarray}
where inequality a) follows by the data-processing inequality and the
convexity of $R_W(\cdot)$. Inequality b) follows by the
data-processing inequality, and c) follows since ${\bf S}_2$ and ${\bf
  S}_1-\rho {\bf S}_2$ are independent. Substituting
$R_W(D_W(\varphi^{(n)}))$ on the LHS of \eqref{eq:mut-inf-Dw} by its
explicit form gives
\begin{IEEEeqnarray*}{rCl}
\frac{n}{2} \log_2 \left(
  \frac{\sigma^2(1-\rho^2)}{D_W(\varphi^{(n)})} \right) & \leq & I({\bf
  S}_1;{\bf Y}|{\bf S}_2).
\end{IEEEeqnarray*}
Rewriting this inequality establishes \eqref{eq:mut-inf-bd}.

Lemma \ref{lemma:mut-info} and Inequality \eqref{eq:mut-inf-bd}
combine to prove \eqref{eq:lb-Dw}. We next derive an upper bound on
$D_W(\varphi^{(n)})$.

\subsection{``Upper Bound'' on minimal $D_W(\varphi^{(n)})$}

We now present an estimator $\tilde{\varphi}^{(n)}({\bf S}_2,{\bf Y})$
for which we show that 
\begin{IEEEeqnarray}{rCl}\label{eq:ub-Dw}
  \text{Assumption \ref{as:false} } & \quad \Rightarrow \quad &
  \varlimsup_{\nu \rightarrow \infty} D_W(\tilde{\varphi}^{(n_{\nu})}) <
  \sigma^2 (1-\rho^2) \frac{N}{N + P_1(1-\rho^2)}, \hspace{5mm}
\end{IEEEeqnarray}
for some monotonically increasing sequence $\{ n_{\nu} \}$ of
integers. From Implications \eqref{eq:ub-Dw} and \eqref{eq:lb-Dw} we
then conclude that Assumption \ref{as:false} is false. The estimator
$\tilde{\varphi}^{(n)}({\bf S}_2,{\bf Y})$ is given by
\begin{IEEEeqnarray*}{rCl}\label{eq:macfb-phi-tilde}
\tilde{\varphi}^{(n)}({\bf S}_2,{\bf Y}) & \triangleq & \alpha
\hat{\bf S}_1 - \beta {\bf S}_2 \nonumber\\
& = & \alpha \E{{\bf S}_1|{\bf Y}} - \beta {\bf S}_2,
\end{IEEEeqnarray*}
where the coefficients $\alpha$ and $\beta$ are given by
\begin{IEEEeqnarray}{rCl}
  \alpha & \triangleq & \frac{\sigma^2\left( \sqrt{\sigma^2-D_1^{\ast}} -
      \rho\sqrt{\sigma^2-D_2^{\ast}} \right)}{D_2^{\ast}
    \sqrt{\sigma^2-D_1^{\ast}}} \label{eq:macfb-prf-uc-alpha}\\[4mm]
  \beta & \triangleq &
  \frac{\sqrt{(\sigma^2-D_1^{\ast})(\sigma^2-D_2^{\ast})} -\rho
    (\sigma^2-D_2^{\ast})}{D_2^{\ast}}, \label{eq:macfb-prf-uc-beta}
\end{IEEEeqnarray}
with $(D_1^{\ast},D_2^{\ast})$ as in Assumption \ref{as:false}. The
idea for showing that for this estimator \eqref{eq:ub-Dw} holds, is to
exploit the fact that if ${\bf Y}$ allows for a ``good'' estimate
$\hat{\bf S}_1$ of ${\bf S}_1$, i.e.~if $D_1^{\ast} <
D_1^{\textnormal{u}}$, then Transmitter 2 can also make a ``good''
estimate of ${\bf W}$, based on ${\bf S}_2$ and ${\bf Y}$. To show
this we first notice that Assumption \ref{as:false} implies that there
exists a monotonically increasing sequence of integers $\{ n_{\nu} \}$
such that
\begin{IEEEeqnarray}{rCl}\label{eq:macfb-rdc2}
  \lim_{\nu \rightarrow \infty} \frac{1}{n_{\nu}} \sum_{k=1}^{n_{\nu}}
  \E{(S_{i,k} - \hat{S}_{i,k})^2} & = & D_i^{\ast} \qquad i \in \{
  1,2 \}.
\end{IEEEeqnarray}
We now derive \eqref{eq:ub-Dw} using the following two lemmas.

\begin{lm}\label{lemma:expectation-bound}
  For every $\delta > 0$ there exists an $\nu_0(\delta)$ such that for
  all $\nu > \nu_0(\delta)$ the following inequalities hold
  \begin{IEEEeqnarray}{rCl}
    \frac{1}{n_{\nu}} \sum_{k=1}^{n_{\nu}} \E{S_{1,k} \hat{S}_{1,k}} &
    \geq & \sigma^2 - D^{\ast}_1 - \delta, \label{eq:corr-bound-S1-S1h} \\
    \frac{1}{n_{\nu}} \sum_{k=1}^{n_{\nu}} \E{\hat{S}_{1,k}^2} & \leq
    & \sigma^2-D^{\ast}_1 + \delta, \label{eq:corr-bound-S1h2}\\
    \frac{1}{n_{\nu}} \sum_{k=1}^{n_{\nu}}\E{\hat{S}_{1,k} S_{2,k}} &
    \leq & \sqrt{(\sigma^2-D_1^{\ast})(\sigma^2-D_2^{\ast}) +
      \delta(\sigma^2+\delta)}. \label{eq:corr-bound-S2-S1h}
  \end{IEEEeqnarray}
\end{lm}

\begin{proof}
  See Appendix \ref{apx:prf-lm2-as}.
\end{proof}

\begin{lm}\label{lemma:non-neg-coeff}
  Assumption \ref{as:false} and in particular $(D_1^{\ast},D_2^{\ast})
  \in \mathscr{D}_3$ implies that the coefficients $\alpha$ and
  $\beta$ defined in \eqref{eq:macfb-prf-uc-alpha} and
  \eqref{eq:macfb-prf-uc-beta} satisfy
  \begin{IEEEeqnarray}{rCl}\label{eq:non-neg-coeff}
    \alpha \geq 0 & \qquad \text{and} \qquad & (\rho - \beta) \geq 0.
  \end{IEEEeqnarray}
\end{lm}

\begin{proof}
  Follows by noting that for every $(D_1^{\ast},D_2^{\ast}) \in
  \mathscr{D}_3$
  \begin{IEEEeqnarray*}{rCll}
    \hspace{18mm} D_2^{\ast} & \geq & \left\{ \begin{array}{l l}
        \left( \sigma^2 (1-\rho^2) -D_1^{\ast}\right)
        \frac{\sigma^2}{\sigma^2-D_1^{\ast}} & \text{if } 0 \leq
        D_1^{\ast} \leq \sigma^2(1-\rho^2),\\[2mm]
        \frac{\left( D_1^{\ast} - \sigma^2 (1-\rho^2)
          \right)}{\rho^2} & \text{if } D_1^{\ast} >
        \sigma^2(1-\rho^2).
      \end{array} \right. \hspace{18mm} & \\[-5mm]
    & & & \qedhere
  \end{IEEEeqnarray*}
\end{proof}

Using Lemma \ref{lemma:expectation-bound} and Lemma
\ref{lemma:non-neg-coeff} we now prove \eqref{eq:ub-Dw} as follows:
\begin{IEEEeqnarray}{rCl}
  D_W(\tilde{\varphi}^{(n_{\nu})}) & = & \frac{1}{n_{\nu}} \E{ \| {\bf W} -
    \tilde{\varphi}({\bf S}_2,{\bf Y}) \|^2} \nonumber\\[2mm]
  & = & \frac{1}{n_{\nu}}
  \sum_{k=1}^{n_{\nu}} \E{(S_{1,k}-\rho S_{2,k} -\alpha \hat{S}_{1,k}
    + \beta S_{2,k})^2} \nonumber\\[2mm]
  & = & \frac{1}{n_{\nu}} \sum_{k=1}^{n_{\nu}} \E{(S_{1,k} -\alpha
    \hat{S}_{1,k} - (\rho - \beta ) S_{2,k})^2} \nonumber\\[2mm]
  & = & \frac{1}{n_{\nu}} \sum_{k=1}^{n_{\nu}} \bigg( \E{S_{1,k}^2} -
  2 \alpha \E{S_{1,k} \hat{S}_{1,k}} - 2(\rho-\beta ) \E{S_{1,k}
    S_{2,k}} \nonumber\\ 
  & & \hspace{13mm} {} + \alpha^2 \E{\hat{S}_{1,k}^2}  + 2\alpha (\rho
  - \beta) \E{\hat{S}_{1,k} S_{2,k}} \nonumber\\
  & & \hspace{13mm} {} + (\rho-\beta)^2 \E{S_{2,k}^2}
  \bigg). \label{eq:achv-Dw-part1}
\end{IEEEeqnarray}
Using Lemma \ref{lemma:expectation-bound} and Lemma
\ref{lemma:non-neg-coeff}, as well as $\E{S_{1,k}} = \E{S_{2,k}} =
\sigma^2$ and $\E{S_{1,k}S_{2,k}} = \rho \sigma^2$, we now get that
for $P_1$, $P_2$, $N$ satisfying \eqref{eq:mac-SNRcond-uc-opt} and for
every $\delta > 0$ there exists an $\nu_0(\delta) > 0$ such that for
all $\nu > \nu_0(\delta)$,
\begin{IEEEeqnarray}{rCl}
  D_W(\tilde{\varphi}^{(n_{\nu})}) & \leq &
  \sigma^2 - 2 \alpha (\sigma^2 - D_1^{\ast} - \delta) -
  2(\rho-\beta ) \rho \sigma^2 \nonumber\\
  & & + \alpha^2 (\sigma^2 - D_1^{\ast} + \delta) + 2\alpha (\rho
  - \beta) \left(\sqrt{(\sigma^2-D_1^{\ast})(\sigma^2-D_2^{\ast}) +
      \delta(\sigma^2+\delta)} \right) \nonumber\\
  & & + (\rho-\beta)^2 \sigma^2. \label{eq:achv-Dw-part2}
\end{IEEEeqnarray}
Letting $\nu$ tend to infinity and then $\delta \rightarrow 0$ we
obtain from \eqref{eq:achv-Dw-part2} that
\begin{IEEEeqnarray}{rCl}
\varlimsup_{\nu \rightarrow \infty} D_W(\tilde{\varphi}^{(n_{\nu})})
& \leq & \sigma^2 - 2 \alpha (\sigma^2 - D_1^{\ast}) -
2(\rho-\beta ) \rho \sigma^2 \nonumber\\
& & + \alpha^2 (\sigma^2 - D_1^{\ast}) + 2\alpha (\rho
- \beta) \sqrt{(\sigma^2 - D_1^{\ast})(\sigma^2 - D_2^{\ast})} \nonumber\\
& & + (\rho-\beta)^2 \sigma^2 \nonumber\\[5mm]
& = & \sigma^2 \; \frac{\; 2\rho
  \sqrt{(\sigma^2-D_1^{\ast})(\sigma^2-D_2^{\ast})}+
  D_1^{\ast}+D_2^{\ast}-\sigma^2(1+\rho^2)}{D_2^{\ast}}, \label{eq:macfb-ub1-Dw}
\end{IEEEeqnarray}
where in the last step we have replaced the terms $\alpha$ and $\beta$
by their expressions in \eqref{eq:macfb-prf-uc-alpha} and
\eqref{eq:macfb-prf-uc-beta}. To conclude our upper bound we now make
use of one last lemma.

\begin{lm}\label{lm:macfb-decr-ub-Dw}
  For all $(D_1^{\ast},D_2^{\ast}) \in
  \textnormal{int}(\mathscr{D}_3)$ the expression on the RHS of
  \eqref{eq:macfb-ub1-Dw} is strictly increasing in $D_1^{\ast}$.
\end{lm}

\begin{proof}
  Denote by $\tilde{D}_W$ the RHS of \eqref{eq:macfb-ub1-Dw}.  The
  proof follows by showing that
  for all $(D_1^{\ast},D_2^{\ast}) \in \textnormal{int}(\mathscr{D}_3)$
  \begin{IEEEeqnarray*}{rCl}
    \frac{\partial \tilde{D}_W}{\partial D_1^{\ast}} & > & 0.
  \end{IEEEeqnarray*}
  This follows by direct differentiation and by noting that for
  $(D_1^{\ast},D_2^{\ast}) \in \textnormal{int}(\mathscr{D}_3)$ 
  \begin{IEEEeqnarray*}{rCll}
    \hspace{18mm} D_2^{\ast} & > & \left\{ \begin{array}{l l}
        \left( \sigma^2 (1-\rho^2) -D_1^{\ast}\right)
        \frac{\sigma^2}{\sigma^2-D_1^{\ast}} & \text{if } 0 \leq
        D_1^{\ast} \leq \sigma^2(1-\rho^2),\\[2mm]
        \frac{\left( D_1^{\ast} - \sigma^2 (1-\rho^2)
          \right)}{\rho^2} & \text{if } D_1^{\ast} >
        \sigma^2(1-\rho^2).
      \end{array} \right. \hspace{18mm} & \\[-5mm]
    & & & \qedhere
  \end{IEEEeqnarray*}
  %
  %
  
\end{proof}

Since $D_1^{\ast} < D_1^{\textnormal{u}}$ and $D_2^{\ast} =
D_2^{\textnormal{u}}$ it follows from \eqref{eq:macfb-ub1-Dw} and
Lemma \ref{lm:macfb-decr-ub-Dw} that
\begin{IEEEeqnarray}{rCl}
\varlimsup_{\nu \rightarrow \infty} D_W(\tilde{\varphi}^{(n_{\nu})})
& < & \sigma^2 \; \frac{\; 2\rho
  \sqrt{(\sigma^2-D_1^{\textnormal{u}})(\sigma^2-D_2^{\textnormal{u}})}+
  D_1^{\textnormal{u}} + D_2^{\textnormal{u}} -
  \sigma^2(1+\rho^2)}{D_2^{\textnormal{u}}}  \nonumber\\[2mm]
& = & \sigma^2 \frac{N(1-\rho^2)}{P_1(1-\rho^2)+N},
\end{IEEEeqnarray}
where the last line follows from replacing $D_1^{\textnormal{u}}$ and
$D_2^{\textnormal{u}}$ by their expressions given in
Theorem~\ref{thm:macfb-uncoded}. Thus, we have proven \eqref{eq:ub-Dw}.


\subsection{Concluding the Proof of Theorem
  \ref{thm:macfb-uncoded}} \label{sec:macfb-uc-prf-conclude}

It follows from \eqref{eq:lb-Dw} and \eqref{eq:ub-Dw} that Assumption
\ref{as:false} is false.
We now show that this implies that if $P_1$, $P_2$, $N$ satisfy
\eqref{eq:mac-SNRcond-uc-opt} with strict inequality, then no pair
$(D_1,D_2)$ satisfying $D_1 < D_1^{\textnormal{u}}$ and $D_2 \leq
D_2^{\textnormal{u}}$ or satisfying $D_1 \leq D_1^{\textnormal{u}}$
and $D_2 < D_2^{\textnormal{u}}$ is achievable. To prove this we
assume $\rho > 0$ because for $\rho = 0$ Condition
\eqref{eq:mac-SNRcond-uc-opt} becomes $P_1 P_2 \leq 0$ and is
therefore never satisfied with strict inequality.

Our arguments are given in the following sequence of statements:\\[2mm]
\begin{itemize}
\item[A)] If $P_1$, $P_2$, $N$ satisfy \eqref{eq:mac-SNRcond-uc-opt}
  with strict inequality, then the set of $(D_1^{\ast},D_2^{\ast})$
  satisfying \eqref{eq:macfb-ass-contr} is not empty.\\[-1mm]
\end{itemize}
\hfill \parbox{14.1cm}{ Statement A) holds since if $P_1$, $P_2$, $N$
  satisfy \eqref{eq:mac-SNRcond-uc-opt} with strict inequality, then
  $(D_1^{\textnormal{u}},
  D_2^{\textnormal{u}})~\in~\textnormal{int}(\mathscr{D}_3)$ and
  $\textnormal{int}(\mathscr{D}_3) \neq \emptyset$ whenever $\rho \neq
  0$.}\\[6mm]
\begin{itemize}
\item[B)] If $P_1$, $P_2$, $N$ satisfy \eqref{eq:mac-SNRcond-uc-opt}
  with strict inequality, then there do not exist encoding rules, that,
  when combined with the optimal conditional expectation
  reconstructors, result in $(D_1^{\ast},D_2^{\ast})$ as defined in
  \eqref{eq:macfb-def-Di*} satisfying
  \begin{IEEEeqnarray*}{rCl}
    D_1^{\ast} < D_1^{\textnormal{u}} & \quad \qquad \text{and} \quad
    \qquad & D_2^{\ast} = D_2^{\textnormal{u}},
  \end{IEEEeqnarray*}
  (with $(D_1^{\ast},D_2^{\ast})$ in or outside
  $\textnormal{int}(\mathscr{D}_3)$).\\[1mm]
\end{itemize}
\hfill \parbox{14.1cm}{ Statement B) can be shown by
  contradiction. If a coding scheme as described
  in B) were to exist, then by time-sharing it with the
  uncoded scheme---for which $(D_1^{\textnormal{u}},
  D_2^{\textnormal{u}}) \in \textnormal{int}(\mathscr{D}_3)$---and
  by Statement A), we would obtain a scheme for which
  $(D_1^{\ast},D_2^{\ast})$ satisfies \eqref{eq:macfb-ass-contr}, in
  contradiction to the fact that Assumption \ref{as:false} is false.}

\begin{itemize}
\item[C)] If $P_1$, $P_2$, $N$ satisfy \eqref{eq:mac-SNRcond-uc-opt}
  with strict inequality, then there exist no encoding rules, which,
  when combined with the optimal conditional expectation
  reconstructors, result in $(D_1^{\ast},D_2^{\ast})$ as defined in
  \eqref{eq:macfb-def-Di*} such that
  \begin{IEEEeqnarray*}{rCl}
    D_1^{\ast} = D_1^{\textnormal{u}} & \quad \qquad \text{and} \quad
    \qquad & D_2^{\ast} < D_2^{\textnormal{u}}.\\[-3mm]
  \end{IEEEeqnarray*}
\end{itemize}
\hfill \parbox{14.1cm}{ Statement C) can be proved using arguments
  similar to those used to prove Statement~B).}\\[6mm]

\begin{itemize}
\item[D)] If $P_1$, $P_2$, $N$ satisfy \eqref{eq:mac-SNRcond-uc-opt}
  with strict inequality, then there exist no encoding rules, which
  when combined with the optimal conditional expectation
  reconstructors, result in $(D_1^{\ast},D_2^{\ast})$ as defined in
  \eqref{eq:macfb-def-Di*} such that $D_1^{\ast} <
  D_1^{\textnormal{u}}$ and $D_2^{\ast} \leq D_2^{\textnormal{u}}$ or
  such that $D_1^{\ast} \leq D_1^{\textnormal{u}}$ and $D_2^{\ast} <
  D_2^{\textnormal{u}}$.\\[3mm]
\end{itemize}
\hfill \parbox{14.1cm}{ To show Statement D) we proceed by
  contradiction. To this end, consider two variations of our uncoded
  scheme. Call these two variations ``Scheme U$_1$'' and ``Scheme
  U$_2$''. Let Scheme U$_1$ be given by the channel inputs
  \begin{IEEEeqnarray*}{rCl}
    X_{1,k}^{\textnormal{u}_1} = \sqrt{\frac{P_1}{\sigma^2}} S_{1,k} & \qquad \text{and}
    \qquad & X_{2,k}^{\textnormal{u}_1} = 0,
  \end{IEEEeqnarray*}
  and the optimal conditional expectation reconstructors $\hat{\bf
    S}_1 = \E{{\bf S}_1 \big| {\bf Y}}$ and $\hat{\bf S}_2 = \E{{\bf
      S}_2 \big| {\bf Y}}$. The resulting distorion pair
  $(D_1^{\textnormal{u}_1},D_2^{\textnormal{u}_1})$ is given by
  \begin{IEEEeqnarray*}{rCl}
    D_1^{\textnormal{u}_1} = \sigma^2 \frac{N}{P_1+N} & \qquad
    \qquad & D_2^{\textnormal{u}_1} = \sigma^2
    \frac{(1-\rho^2)P_1 +N}{P_1+N}.
  \end{IEEEeqnarray*}
  Similarly, let Scheme U$_2$ be given by the channel inputs
  \begin{IEEEeqnarray*}{rCl}
    X_{1,k}^{\textnormal{u}_2} = 0 & \qquad \text{and} \qquad &
    X_{2,k}^{\textnormal{u}_2} = \sqrt{\frac{P_2}{\sigma^2}} S_{2,k},
  \end{IEEEeqnarray*}
  and the same optimal conditional expectation reconstructors as for
  Scheme U$_1$. The resulting distorion pair
  $(D_1^{\textnormal{u}_2},D_2^{\textnormal{u}_2})$ is given by
  \begin{IEEEeqnarray*}{rCl}
    D_1^{\textnormal{u}_2} = \sigma^2 \frac{(1-\rho^2)P_2
      +N}{P_2+N} & \qquad \qquad & D_2^{\textnormal{u}_2} =
    \sigma^2 \frac{N}{P_2+N}.
  \end{IEEEeqnarray*}
  Now assume there would exist a coding scheme as described in D).
  Since $D_2^{\textnormal{u}_1} > D_2^{\textnormal{u}}$ and
  $D_1^{\textnormal{u}_2} > D_1^{\textnormal{u}}$ it would follow from
  time-sharing either with Scheme U$_1$ or Scheme U$_2$
  that Statement B) or Statement C) is false.}\\[6mm]
\begin{itemize}
\item[E)] If $P_1$, $P_2$, $N$ satisfy \eqref{eq:mac-SNRcond-uc-opt}
  with strict inequality, then there exist no coding scheme resulting
  in $(D_1^{\ast},D_2^{\ast})$ as defined in \eqref{eq:macfb-def-Di*}
  such that
  \begin{IEEEeqnarray*}{rCl}
    D_1^{\ast} < D_1^{\textnormal{u}} & \quad \qquad \text{and} \quad
    \qquad & D_2^{\ast} \leq D_2^{\textnormal{u}},
  \end{IEEEeqnarray*}
  (be the reconstruction rule optimal or not).\\[1mm]
\end{itemize}
\hfill \parbox{14.1cm}{ Statement E) follows from D) because no
  reconstructor $\phi_i^{(n)}$ can outperform the optimal conditional
  expectation reconstructor $\hat{\bf S}_i = \E{ {\bf S}_i \big| {\bf
      Y}}$.}\\[7mm]
By Statement E) it follows that if $P_1$, $P_2$, $N$ satisfy
\eqref{eq:mac-SNRcond-uc-opt} with strict inequality, then no
$(D_1,D_2)$ satisfying $D_1 < D_1^{\textnormal{u}}$ and $D_2 \leq
D_2^{\textnormal{u}}$ is achievable.


\subsection{Proof of Lemma \ref{lemma:expectation-bound}} \label{apx:prf-lm2-as}

By \eqref{eq:macfb-rdc2} it follows that for every $\delta > 0$ there
exists a $\nu_0(\delta) > 0$ such that for all $\nu > \nu_0(\delta)$
\begin{IEEEeqnarray}{rCl}\label{eq:distortion}
D^{\ast}_i - \delta < & \frac{1}{n_{\nu}} \sum_{k=1}^{n_{\nu}} \E{(S_{i,k} -
  \hat{S}_{i,k})^2} < & D^{\ast}_i + \delta \qquad i \in \{ 1,2 \}.
\end{IEEEeqnarray}
Using \eqref{eq:distortion}, the relation $\E{S_{1,k}^2} = \sigma^2$,
and \eqref{eq:macfb-hatSi} we obtain that
\begin{IEEEeqnarray}{rCl}\label{eq:bound-corr-S1S1h}
\sigma^2 - D^{\ast}_1 -\delta \leq & \frac{1}{n_{\nu}} \sum_{k=1}^{n_{\nu}}
\E{S_{1,k}\hat{S}_{1,k}} & \leq \sigma^2 - D^{\ast}_1 +\delta,
\end{IEEEeqnarray}
and that
\begin{IEEEeqnarray}{rCl}\label{eq:bound-corr-S1h2}
\sigma^2 - D^{\ast}_1 -\delta \leq & \frac{1}{n_{\nu}} \sum_{k=1}^{n_{\nu}}
\E{\hat{S}_{1,k}^2} & \leq \sigma^2 - D^{\ast}_1 + \delta.
\end{IEEEeqnarray}
This proves Inequalities \eqref{eq:corr-bound-S1-S1h} and
\eqref{eq:corr-bound-S1h2}.

To prove \eqref{eq:corr-bound-S2-S1h} we note that for every $c \in
\Reals$ we can view $c \hat{S}_{1,k}$ as an estimator of $S_{2,k}$
based on ${\bf Y}$. As such it cannot outperform the optimal estimator
of $S_{2,k}$ given by ${\bf Y}$, namely the estimator $\hat{\bf S}_2 =
\E{{\bf S}_2 | {\bf Y} }$. Consequently, for every $\delta > 0$ it
follows by \eqref{eq:distortion} that there exists an
$\nu_0(\delta)>0$ such that for all $\nu > \nu_0(\delta)$ and all $c \in
\Reals$,
\begin{IEEEeqnarray}{rCl}\label{eq:prf-macfb-innerS2S1h}
  \frac{1}{n_{\nu}} \sum_{k=1}^{n_{\nu}} \E{(S_{2,k} - c
    \hat{S}_{1,k})^2} & \geq & \frac{1}{n_{\nu}} \sum_{k=1}^{n_{\nu}}
  \E{(S_{2,k} - \hat{S}_{1,k})^2} \nonumber\\[3mm]
  & > & D_2^{\ast} - \delta.
\end{IEEEeqnarray}
Rewriting \eqref{eq:prf-macfb-innerS2S1h} gives
\begin{IEEEeqnarray*}{rCl}
  \sigma^2 - 2c \frac{1}{n_{\nu}} \sum_{k=1}^{n_{\nu}} \E{S_{2,k}\hat{S}_{1,k}} +
  c^2 (\sigma^2 - D_1^{\ast} + \delta) > D_2^{\ast} - \delta,
\end{IEEEeqnarray*}
and choosing
\begin{IEEEeqnarray*}{rCl}
  c & = & \sqrt{\frac{\sigma^2-D_2^{\ast}-\delta}{\sigma^2-D_1^{\ast}+\delta}},
\end{IEEEeqnarray*}
yields that for all $\nu > \nu_0(\delta)$
\begin{IEEEeqnarray*}{rCll}
  \hspace{17mm} \frac{1}{n_{\nu}} \sum_{k=1}^{n_{\nu}} \E{S_{2,k}\hat{S}_{1,k}} & \leq &
  \sqrt{(\sigma^2-D_1^{\ast}+\delta)(\sigma^2-D_2^{\ast}-\delta)} & \\
  & = & \sqrt{(\sigma^2-D_1^{\ast})(\sigma^2-D_2^{\ast}) -
    \delta(D_2^{\ast}-D_1^{\ast}+\delta)} & \\[3mm]
  & \leq & \sqrt{(\sigma^2-D_1^{\ast})(\sigma^2-D_2^{\ast}) +
    \delta(\sigma^2+\delta)}. & \hspace{18mm} \qed
\end{IEEEeqnarray*}

\end{document}